\documentclass[onecolumn,pre,floats,aps,amsmath,amssymb,nofootinbib]{revtex4-2}
\usepackage{comment}
\usepackage[utf8]{inputenc}
\usepackage{mathtools}
\usepackage{bbm}
\usepackage{cancel} 
\usepackage{dcolumn}% Align table columns on decimal point
\usepackage{blindtext}
\usepackage{xcolor}
\usepackage{amsthm}
\usepackage{bm}% bold math
\usepackage{hyperref}% add hypertext capabilities
\usepackage{graphicx}
\usepackage{graphics,epsfig}
\usepackage{MnSymbol}
\usepackage{subcaption}
\usepackage{verbatim}
\usepackage{microtype}
\usepackage{adjustbox}
\usepackage{slashed}
\usepackage{physics}
\usepackage{soul}
\newcommand{\be}{\begin{equation}}
\newcommand{\ee}{\end{equation}}
\newcommand{\bea}{\begin{eqnarray}}
\newcommand{\eea}{\end{eqnarray}}
\newcommand{\bs}{\boldsymbol}
\newcommand{\beq}{\begin{equation}}
\newcommand{\eeq}{\end{equation}}

\newcommand{\nn}{\nonumber}

 \newcommand{\pslash}{{\not{\hspace{-0.08cm}p}}}

\newcommand{\Romatre}{Dipartimento di Matematica e Fisica, Universit\`a  Roma Tre and INFN, Sezione di Roma Tre,\\ Via della Vasca Navale 84, I-00146 Rome, Italy}
\newcommand{\RomatreINFN}{Istituto Nazionale di Fisica Nucleare, Sezione di Roma Tre,\\ Via della Vasca Navale 84, I-00146 Rome, Italy}
\newcommand{\Romadue}{Dipartimento di Fisica and INFN, Universit\`a di Roma ``Tor Vergata",\\ Via della Ricerca Scientifica 1, I-00133 Roma, Italy}

\begin{document}
\title{Inclusive hadronic decay rate of the \texorpdfstring{$\tau$}{tau} lepton from lattice QCD} 
\author{A.\,Evangelista}\affiliation{\Romadue}
\author{R.\,Frezzotti}\affiliation{\Romadue} 
\author{G.\,Gagliardi}\affiliation{\RomatreINFN}
\author{V.\,Lubicz}\affiliation{\Romatre} 
\author{F.\,Sanfilippo}\affiliation{\RomatreINFN}
\author{S.\,Simula}\affiliation{\RomatreINFN}
\author{N.\,Tantalo}\affiliation{\Romadue}

\date{\today}

\setlength{\parskip}{14pt}
\setlength{\parindent}{0pt}

\begin{abstract}
\vspace{0.05cm}
\centerline{\includegraphics[height=4.7cm]{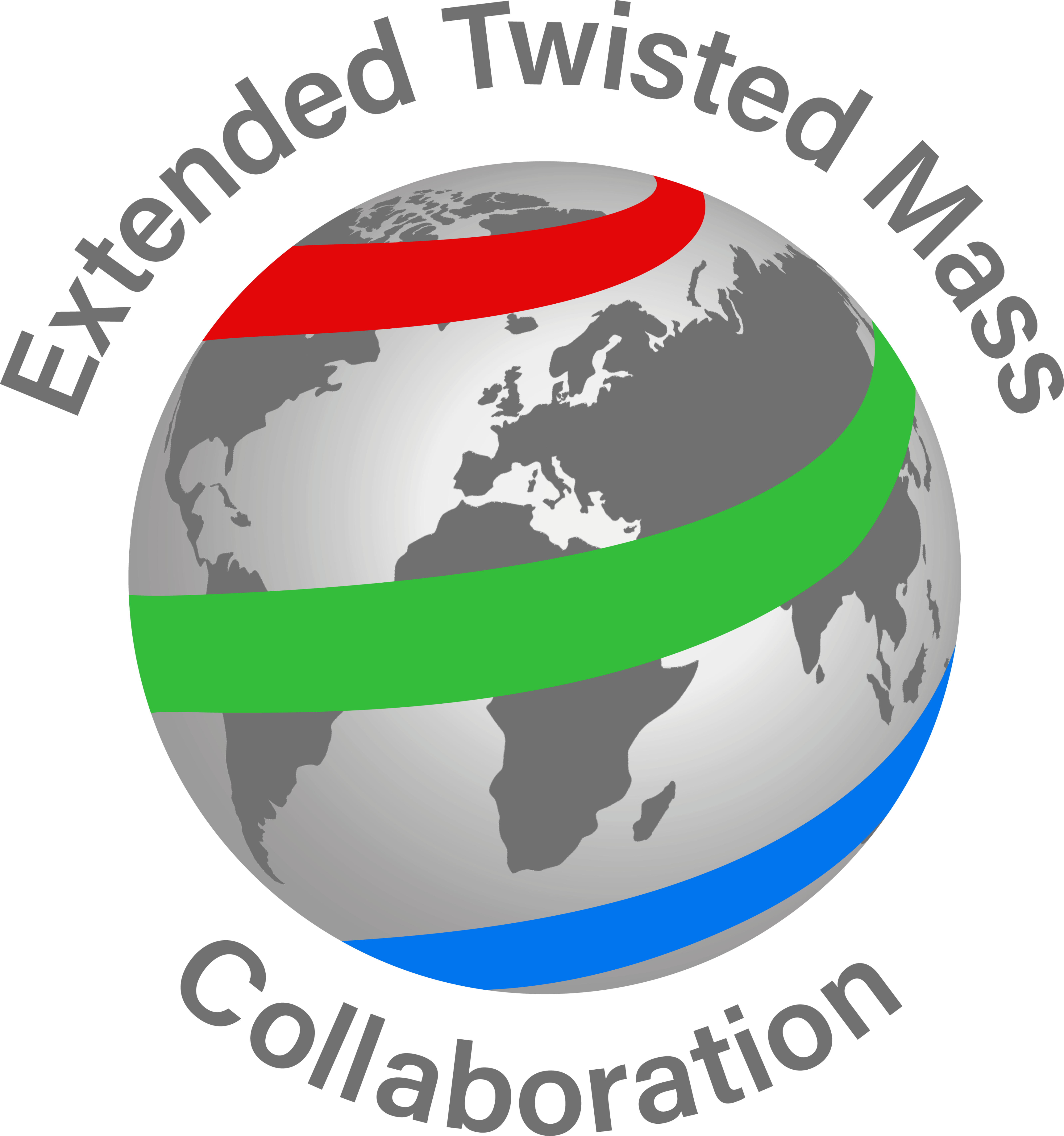}}
\vspace{0.3cm}
Inclusive hadronic decays of the $\tau$ lepton are very interesting from the phenomenological point of view since they give access to the CKM matrix elements $V_{ud}$ and $V_{us}$. In this paper, for the first time, by employing the HLT method for hadronic smeared spectral densities we compute on the lattice the inclusive decay rate of the processes $\tau  \to X_{ud}\, \nu_\tau$, where $X_{ud}$ is a generic hadronic state with $\bar u d$ flavor quantum numbers.
Our computation, which avoids any recourse to OPE and/or perturbative approximations, is carried out in isospin symmetric $N_{f}=2+1+1$ lattice QCD at physical quark masses, using ensembles produced by the ETMC at three lattice spacings and two volumes. All uncertainties, except for isospin breaking effects, are taken into account and a result with a subpercent error is obtained for $|V_{ud}|$, which is nicely consistent with the current world average. These findings validate our approach and also motivate the inclusion of isospin breaking corrections and its extension to the inclusive decay $\tau \to X_{us}\,  \nu_\tau$, paving the way towards a high-precision first principles determination of $|V_{us}|$ and $|V_{ud}|$ from inclusive $\tau$ decay.
\end{abstract}

\maketitle

\section{Introduction}

The $\tau$ lepton, owing to its large mass, is the only lepton that can decay through weak interactions in both hadrons and leptons. From the phenomenological point of view, hadronic $\tau$ decays, both inclusive ($\tau \to X_{ud/us}\, \nu_\tau$) and exclusive ($\tau \to \pi/K \,\nu_\tau$), are very interesting, since they give access to the CKM matrix elements $|V_{ud}|$ and $|V_{us}|$, thus providing additional information on these parameters besides the one coming from the currently more precise determinations of $|V_{ud}|$ from nuclear super-allowed beta transitions and $|V_{us}|$ and $|V_{us}|/|V_{ud}|$ from $K$ and $\pi$ leptonic and semileptonic decays (see the reviews FLAG21~\cite{FlavourLatticeAveragingGroupFLAG:2021npn}
and HFLAV22~\cite{HeavyFlavorAveragingGroup:2022wzx}).

The phenomenological studies of inclusive hadronic $\tau$ decays have been mainly focused so far on the strange-hadronic decays and the determination of $|V_{us}|$. A direct OPE analysis of the inclusive process $\tau \to X_{us}\, \nu_\tau$ leads to the result
%\be \label{stdincVus}
$|V_{us}|_{\rm \tau-incl-1} = 0.2184(21)$~\cite{Gamiz:2006xx,Pich:2013lsa}.
%\ee
This determination, however, is in disagreement with both the value of $|V_{us}|$ obtained from the analysis of leptonic and semileptonic kaon decays, namely
%\be \label{Kl23Vus}
$|V_{us}|_{\rm K\ell 3 - K/\pi\ell 2} = 0.2248(6)$~\cite{FlavourLatticeAveragingGroupFLAG:2021npn},
%\ee
and the one obtained by combining CKM unitarity with the measurements of $|V_{ud}|$ from superallowed nuclear $\beta$-transitions, namely
%\be \label{uniVus}
$|V_{us}|_{\rm uni} = 0.2277(13)$~\cite{FlavourLatticeAveragingGroupFLAG:2021npn,HeavyFlavorAveragingGroup:2022wzx}.
%\ee
Moreover, the value of $|V_{us}|$ obtained from inclusive $\tau$ decays is also smaller than, though not incompatible with, the one obtained from the analysis of exclusive $\tau \to \pi/K \,\nu_\tau$ decays, namely
$|V_{us}|_{\rm \tau-excl} = 0.2222(17)$~\cite{HeavyFlavorAveragingGroup:2022wzx}.
%\ee

The apparent tension between the value of $|V_{us}|$ coming from
inclusive $\tau$ decays and the other determinations, has been critically addressed in Ref.~\cite{Hudspith:2017vew}, where, owing to a new treatment of higher order terms in the OPE, which are determined by fits to lattice current-current correlators, and by using a partly different experimental input, a larger value of $|V_{us}|$ is obtained, namely (according to the update given in 
Ref.~\cite{Maltman:2019xeh}) 
%\be \label{revincVus}
$|V_{us}|_{\rm \tau-incl-2} = 0.2219(22)$.

Recently, a new method to determine $|V_{us}|$ from inclusive $\tau$ decays has been proposed in Ref.~\cite{RBC:2018uyk}. By introducing generalized dispersion relations, involving weights with poles chosen so as to suppress the most uncertain contributions coming from large time-like momenta, this method evaluates the corresponding spectral integrals using lattice-QCD data for the hadronic vacuum polarization function,
thereby avoiding assumptions on OPE and related condensates. In this way, a later analysis~\cite{Maltman:2019xeh} yields
%\be \label{partinclVus}
$|V_{us}|_{\rm \tau-part-incl} = 0.2240(18)$, 
%\ee
in agreement with the determinations from kaon (semi)leptonic decays and exclusive $\tau$ decays. As noted in Ref.~\cite{Crivellin:2022rhw}, this result is in line with the fact that, owing to the particular generalized dispersion relation of choice, it mostly relies on the exclusive $\tau \to K \nu_\tau$ data, which represents instead less than $25\% $ of the inclusive strange-hadronic decays (whence the label ``part-incl'').

In this paper we present a novel lattice field theory approach to the study of inclusive hadronic $\tau$ decays which provides a first-principle and fully non-perturbative determination of the inclusive decay rate, in a given fixed flavour hadronic channel. Our method neither relies on OPE assumption and related condensates nor employs generalized dispersion relations, chosen with the aim of reducing the uncertainties due to the use of perturbative approximations. Rather, we employ the presently well-established HLT method~\cite{Hansen:2019idp,Bulava:2021fre} for spectral density reconstruction in order to obtain, directly from the Euclidean lattice correlation functions, the inclusive decay rate, 
expressed in terms of an integral over the energy of the relevant spectral density weighted by a suitably smeared kernel (see Eq.~\ref{eq:R_sigma} below and Ref.~\cite{Gambino:2020crt} where a similar approach has been proposed to study inclusive decays of heavy mesons). Upon working at physical values of the quark masses, taking the infinite volume limit and performing the zero smearing width as well as the continuum extrapolations, 
we obtain a first-principle, fully non-perturbative determination of the inclusive decay rate. A brief account of this new method was already given in Ref.~\cite{Evangelista:2023vtl}.

In this first study, we apply the method to the computation of the inclusive decay rate of $\tau  \to X_{ud}\, \nu_\tau$. For the numerical calculation, we employed the lattice vector and axial two-point correlators with degenerate quark masses, that were produced with high statistics by ETMC for the study of Ref.\,\cite{ExtendedTwistedMass:2022jpw}. By relying on several lattice ensembles produced with $N_f= 2+1+1$ dynamical quark flavours in the isospin symmetric limit of QCD, at three lattice spacings and two volumes, we are able to take into account all statistical and systematic uncertainties, except for the isospin breaking effects, in close analogy to what was done in 
Refs.~\cite{ExtendedTwistedMass:2022jpw,Alexandrou:2022tyn}.

In terms of the ratio $R^{(\tau)}_{ud} = \Gamma(\tau \to X_{ud} \,\nu_\tau)/\Gamma(\tau \to e\, \bar \nu_e \,\nu_\tau)$, we find the theoretical prediction
\be 
 \label{eq:th_val}
\frac{1}{|V_{ud}|^2} \, R^{(\tau)}_{ud} = 3.650 \pm 0.028  \; , 
\ee
a result that has a subpercent total error and looks nicely consistent with the  experimental value~\cite{HeavyFlavorAveragingGroup:2022wzx}
, namely
 \be
 \label{eq:exp_val}
 \frac{1}{|V_{ud}|^2} \, R^{(\tau)}_{ud} = 3.660 \pm 0.008  \; ,
 \ee
obtained using  $|V_{ud}| = 0.97373~(31)$ from nuclear $\beta$-decays~\cite{FlavourLatticeAveragingGroupFLAG:2021npn}. Alternatively, we can compare our theoretical determination in Eq.\,\eqref{eq:th_val} with the experimental value of $R^{(\tau)}_{ud}$ to obtain
\be
\label{eq:ourVUd}
|V_{ud}| = 0.9752~(39)~.
\ee

We should remind that these results are obtained in the isospin symmetric limit of QCD, thus neglecting, in the determination of the $\tau \to X_{ud} \, \nu_\tau$ decay rate, isospin breaking effects. Nevertheless, these results appear to fully validate our approach and motivate us to compute the leading isospin breaking effects~\cite{deDivitiis:2011eh,deDivitiis:2013xla}, as well as to extend our study to the inclusive process $\tau \to X_{us} \,  \nu_\tau$, thus arriving at a high-precision first-principles and completely non-perturbative determination of $|V_{ud}|$ and $|V_{us}|$ from inclusive $\tau$ decay.

The outline of this paper is as follows: in Sec.~\ref{sec:method} we present our strategy to evaluate the $\tau \to X_{ud} \, \nu_\tau$ decay rate using the HLT method. In Sec.~\ref{sec:numerical_results} we present our numerical results and discuss the continuum and infinite-volume extrapolations, as well as the extrapolation to zero smearing width. We briefly comment on the phenomenological implications of our findings, and compare our results with those of existing experiments. Finally in Sec.~\ref{sec:conclusions} we draw our conclusions. In Appendix~\ref{app:A} we provide an alternative (w.r.t. that given in the main text) derivation of the inclusive $\tau$ decay rate formula in Eq.~(\ref{Gammaf}) based on the optical theorem and the Cutkosky rule, while in Appendix~\ref{app:B}, we derive useful expressions for the leading corrections to the zero-width limit of the smeared ratio defined in Eq.~(\ref{eq:R_sigma}).

\section{The method}
\label{sec:method}

\subsection{The inclusive decay rate}
\begin{figure}[t]
    \centering
    \includegraphics[scale=0.4]{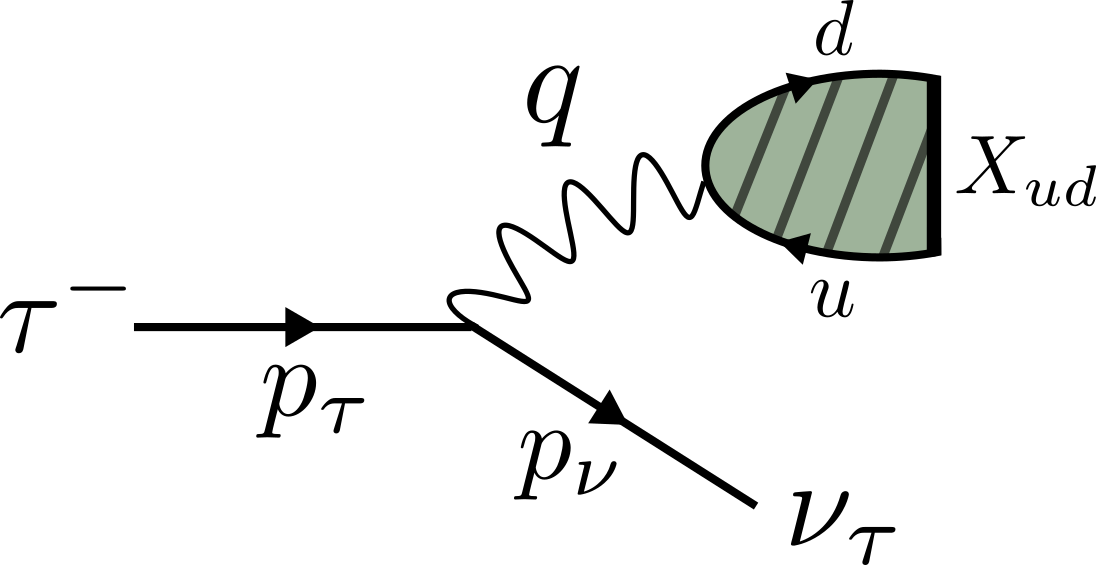}
    \caption{\small\it 
    Feynman diagram representing the amplitude for the $\tau \to X_{ud}\,\nu_{\tau}$ decay.}
    \label{fig:feyn_bis}
\end{figure}

At the leading order in the Fermi effective theory, the amplitude ${\cal A}(\tau \to X_{ud}\, \nu_\tau)$ for the $\tau$-lepton decay into a generic hadronic state $X_{ud}$, with flavor quantum number $\bar u d$, is represented by the Feynman diagram in Fig.\ref{fig:feyn_bis},
%%%%%%%%%%%%%%%%%%%%%%%%%%%%%%%%%%%%%%%%%%%
%%%%%%%%%%%%%%%%%%%%%%%%%%%%%%%%%%%%%%%%%%%
and it is given by
\bea
\label{AX}
    {\cal A}(\tau \to X_{ud}\, \nu_\tau) &=& 
    \frac{G_F}{\sqrt{2}} \, V_{ud}\ \langle \, X_{ud}\, \nu_\tau \, | \,  J_{\nu_\tau\tau }^\alpha (0) J_{ud}^\alpha (0)^\dagger \, | \, \tau \, \rangle = \nn \\
    &=& \frac{G_F}{\sqrt{2}} \, V_{ud}\ \langle \, \nu_\tau \, | \,  J_{\nu_\tau\tau}^\alpha (0) \, | \, \tau \, \rangle \ \langle \, X_{ud} \, | \, J_{ud}^\alpha (0)^\dagger \, | \, 0 \, \rangle~,
\eea
where $J_{\nu_\tau\tau}^\alpha$ and $J_{ud}^\alpha$ are the weak $V-A$ leptonic and hadronic currents 
\be
\label{weak_current}
    J_{\nu_\tau\tau }^\alpha = \bar \nu_\tau \gamma^\alpha (1-\gamma_5) \tau \quad , \quad
    J_{ud}^\alpha = \bar u \gamma^\alpha (1-\gamma_5) d \ .
\ee
The modulus square of the amplitude, summed over the whole set of hadronic final states with $\bar ud $ quantum numbers, is then 
\bea
\label{AXsum}
    |{\cal A}|^2 &=& \sum_X \, |{\cal A}(\tau \to X_{ud}\, \nu_\tau)|^2 = 
%    \nn \\
%    &=& 
     \left(\frac{G_F}{\sqrt{2}}\right)^2 | \, V_{ud}\, |^2 \, L^{\alpha\beta}(p_{\tau}, p_\nu) \, \sum_X \, \langle \, 0  \, | \, J_{ud}^\alpha (0) \, | \, X_{ud}(q) \, \rangle \, \langle \, X_{ud}(q) \, | \, J_{ud}^\beta (0)^\dagger \, | \, 0 \, \rangle    
\eea
where $p_{\tau}$ and $p_\nu$ are the four-momentum of the $\tau$ lepton and of the neutrino, and $q = p_{\tau}-p_\nu$ is the 4-momentum of the final hadronic state. The leptonic tensor $L_{\alpha\beta}(p_{\tau}, p_\nu)$, averaged over the spin orientations of the $\tau$ lepton and summed over those of the $\nu_\tau$ neutrino, is given by
\bea
\label{leptonic}
     L^{\alpha\beta}(p_{\tau}, p_\nu) 
     &=& 
     \frac{1}{2} \Tr\left[ (\, \pslash_{\tau} + m_\tau) \, \gamma^\alpha (1 - \gamma_5) \ \pslash_\nu \gamma^\beta (1 - \gamma_5) \right]  = 
%     \nn \\ 
%     & = & 
     4 \left( p_{\tau}^\alpha p_\nu^\beta + p_{\tau}^\beta p_\nu^\alpha - g^{\alpha\beta} p_{\tau}\cdot p_\nu - i \, \varepsilon^{\alpha\beta\rho\sigma} p_{\tau\rho} p_{\nu\sigma} \right) \ .
\eea
In the hadronic part of the amplitude square, given in Eq.\,\eqref{AXsum}, we now use the identity
\be
\label{complete}
\sum_X \, | \, X(q) \, \rangle \, \langle \, X(q) \, | =  (2\pi)^4 \delta^4 (\mathbb P-q) \ ,
\ee
where $\mathbb P$ is the QCD 4-momentum operator, which expresses the completeness relation in the subspace of states with given 4-momentum $q$. From Eq.\,\eqref{AXsum} one then obtains
\be
    |{\cal A}|^2 = \left(\frac{G_F}{\sqrt{2}}\right)^2 | \, V_{ud}\, |^2 \, L^{\alpha\beta}(p_{\tau}, p_\nu) \, \rho_{ud}^{\alpha\beta}(q) \ .
\ee
where $\rho_{ud}^{\alpha\beta}(q)$ is the spectral density
\be
\label{rho}
    \rho_{ud}^{\alpha\beta}(q) = \langle 0| J_{ud}^\alpha(0)\, (2\pi)^4 \delta^4 (\mathbb P-q) \, J_{ud}^\beta(0)^\dagger |0\rangle \ .
\ee
The (semi-) inclusive decay rate $\Gamma (\tau \to X_{ud}\, \nu_\tau )$ is finally obtained by integrating the modulus square of the amplitude over the phase space and dividing by $2\, m_\tau$ (in the $\tau$-lepton rest frame), where $m_\tau$ is the 
$\tau$-lepton mass:
\bea
\label{Gamma}
    \Gamma_{ud}^{(\tau)} & \equiv & \Gamma (\tau \to X_{ud}\, \nu_\tau ) =
    \frac{1}{2\, m_\tau} \int \frac{d^3 p_\nu}{(2\pi)^3 \, 2 E_\nu} \int \frac{d^4 q}{(2\pi)^4} \ (2\pi)^4\, \delta^4 (p_{\tau}-p_\nu-q) \ |{\cal A}|^2 = \nn \\[0.2cm]
    &=& \frac{G_F^2 \, |V_{ud}|^2}{4\, m_\tau} \int \frac{d^3 p_\nu}{(2\pi)^3 \, 2 E_\nu} \ L_{\alpha\beta}(p_{\tau}, p_\nu) \ \rho_{ud}^{\alpha\beta}(q) \ ,   \qquad q = p_{\tau} - p_\nu \ ,
\eea
where $E_{\nu} = |\bs{p_{\nu}}|$ is the neutrino energy.
In Appendix A we present an alternative, but equivalent, derivation of Eq.\,\eqref{Gamma} based on the optical theorem and the Cutkosky rule.

From Eq.\,\eqref{Gamma} we see that all the long distance effects of the strong interactions in the decay rate are described by the spectral density of Eq.\,\eqref{rho}, whose integral we are going to evaluate non-perturbatively on the lattice. The primary data of the lattice calculation are given by the Euclidean time-dependent correlation function  
\be
\label{Cdit}
    C^{\alpha\beta}(t,\bs{q}) = \int d^3x \, e^{-i \bs{q} \cdot \bs{x}} \, \langle 0 |\, \, T \left( J_{ud}^\alpha(x) \, J_{ud}^\beta(0)^\dagger \right) \, |0\rangle
\ee
at fixed spatial momentum $\bs{q}$.
The relation between the correlator $C^{\alpha\beta}(t,\bs{q})$ and the spectral density $\rho_{ud}^{\alpha\beta}(q)$ is easily derived. By considering $C^{\alpha\beta}(t,\bs{q})$ at positive Euclidean time $t>0$, one has
\bea
\label{Ctt}
C^{\alpha\beta}(t, \bs{q}) &=& \int d^3x \, e^{- i \bs{q} \cdot \bs{x} } \mel{0}{\, J_{ud}^\alpha(x) J_{ud}^\beta(0)^\dagger\,}{0} = \nn \\
&=& \int d^3x \, e^{- i \bs{q} \cdot \bs{x} } \mel{0}{\, J_{ud}^\alpha(0) \, e^{-{\mathbb H}\, t \, +\, i\,\bs{P} \cdot \bs{x}} \, J_{ud}^\beta(0)^\dagger \,}{0}  = \nn \\
&=& \mel{0}{\, J_{ud}^\alpha(0) \, e^{-{\mathbb H}\, t}\, (2\pi)^3 \delta^3(\bs{P}- \bs{q}) \, J_{ud}^\beta(0)^\dagger \,}{0}  = \nn \\
&=& \int_{-\infty}^{+\infty} \frac{dE}{2\pi}\, e^{-E t} \mel{0}{\, J_{ud}^\alpha(0) \, e^{-{\mathbb H}\, t}\, (2\pi)^4 \delta(\mathbb H - E)\, \delta^3(\bs{P}- \bs{q}) \, J_{ud}^\beta(0)^\dagger \,}{0}  = \nn \\
&=& \int_{-\infty}^{+\infty} \frac{dE}{2\pi}\, e^{-E t} \mel{0}{\, J_{ud}^\alpha(0) \, (2\pi)^4 \delta^4({\mathbb P}-q_E)\,  J_{ud}^\beta(0)^\dagger\, }{0} \ ,
\eea
where $q_E \equiv (E, \boldsymbol{q})$ and we have used the identity
\be
\int_{-\infty}^{+\infty} dE \ \delta(\mathbb H - E) = 1 \ .
\ee
The matrix element appearing in Eq.\,\eqref{Ctt} is precisely the spectral density of Eq.\,\eqref{rho}, so that we can write
\be
\label{CtE}
C^{\alpha\beta}(t, \bs{q}) = \int_{0}^{\infty} \frac{dE}{2\pi}\, e^{- E t}\, \rho_{ud}^{\alpha\beta}(E, \bs{q}) \ , \qquad t>0 \ ,
\ee
%where, for later convenience, we have replaced the lower integration limit with $E_0$, where $E_0$ is such that $\rho^{\alpha\beta}(E, \bs{q})=0$ for $E<E_0$. 
Eq.\,\eqref{CtE} provides the relation between the spectral density $\rho_{ud}^{\alpha\beta}(E, \bs{q})$ and the Euclidean correlator $C^{\alpha\beta}(t, \bs{q})$. 

\subsection{Form factors}
Lorentz covariance implies that $\rho_{ud}^{\alpha\beta}(q)$ can be expressed in terms of two scalar form factors, $\rho_{\mathrm{T}}(q^2)$ and $\rho_{\mathrm{L}}(q^2)$, which parameterize respectively the transverse and the longitudinal contribution to the spectral density tensor,  according to
\be
\label{rhoLT}
\rho_{ud}^{\alpha\beta}(q) = \big( q^\alpha q^\beta - q^2\, g^{\alpha\beta} \big) \, \rho_{\mathrm{T}}(q^2) + q^\alpha q^\beta \, \rho_{\mathrm{L}}(q^2) \ .
\ee

We now substitute the spectral density \eqref{rhoLT} and the leptonic tensor \eqref{leptonic} into the expression \eqref{Gamma} for the decay rate. A simple algebra shows that
\be
\label{Lrho}
    L_{\alpha\beta}(p_{\tau}, p_\nu) \ \rho_{ud}^{\alpha\beta}(q) = 2\, m_\tau^4 \, (1-s) \, \big[(1+2\, s)\, \rho_{\mathrm{T}}(s) + \rho_{\mathrm{L}}(s) \big]\ ,
\ee
where 
\be
\label{omega}
s = \frac{q^2}{m_\tau^2} = 1 - 2\, \frac{p_{\tau}\cdot p_\nu}{m_\tau^2} \, ,
\ee
and, to simplify the notation, we have also used $\rho_{\mathrm{T}}(s)\equiv \rho_{\mathrm{T}}(m_\tau^2 s)$.
In the $\tau$-lepton rest frame, $s = 1 - 2\, E_\nu/m_\tau$, so that we can express the integral over the neutrino phase space $\Phi$ in Eq.\,\eqref{Gamma} as an integral over $s$,
\be
\label{phspace}
\int_{\Phi} \frac{d^3 p_\nu}{(2\pi)^3 \, 2 E_\nu} = \frac{1}{4\pi^2} \int_0^{m_\tau/2} d E_\nu \, E_\nu = \frac{m_\tau^2}{16\pi^2} \int_{0}^1 d s \, (1-s) \ .
\ee
Inserting Eqs.\,\eqref{Lrho} and \eqref{phspace} into Eq.\,\eqref{Gamma},  we arrive at the final expression for the inclusive $\tau$ decay rate,
\be
\label{Gammaf}
    \Gamma_{ud}^{(\tau)} =\frac{G_F^2 \, |V_{ud}|^2\, m_\tau^5}{32\pi^2} \int_{0}^1 d s \, (1-s)^2 \, \big[(1+2\, s)\, \rho_{\mathrm{T}}(s) + \rho_{\mathrm{L}}(s)\big] \ .
\ee

In order to compare the theoretical prediction  with the experimental result, we find it convenient to normalize the hadronic rate with the leptonic one, by introducing the ratio
\be
\label{Rtaudef}
 R_{ud}^{(\tau)} = \frac{\Gamma(\tau \to X_{ud} \, \nu_\tau)}{\Gamma(\tau \to e\, \bar \nu_e \, \nu_\tau)} \ .
\ee
A straightforward calculation shows that for the leptonic decay the transverse and longitudinal form factor are given, in the limit of massless electrons and neglecting isospin breaking effects, by
\be 
\rho_{\mathrm{T}}(s) = \frac{1}{3\pi} \quad , \quad \rho_{\mathrm{L}}(s) = 0 \qquad \quad \big(\, \textrm{for} ~ \tau \to e\, \bar \nu_e \, \nu_\tau \big) \, ,
\ee
which in turn leads to the well known result
\be
\label{Gammalep}
    \Gamma(\tau \to e\, \bar \nu_e \, \nu_\tau) =\frac{G_F^2\,  m_\tau^5}{192\, \pi^3} \ .
\ee
Therefore, inserting Eqs.\,\eqref{Gammaf} and \eqref{Gammalep} into Eq.\,\eqref{Rtaudef}, and including a factor $S_{EW} = 1.0201(3)$ to account for the short–distance electroweak correction \cite{Erler:2002mv}, one arrives at the simple result 
\be
\label{Rtau}
     R_{ud}^{(\tau)} = 6 \pi \, S_{EW} \, |V_{ud}|^2 \int_{0}^1 d s \, (1-s)^2 \, \big[(1+2\, s)\, \rho_{\mathrm{T}}(s) + \rho_{\mathrm{L}}(s)\big] \ .
\ee
which expresses the ratio $R_{ud}^{(\tau)}$ in terms of the CKM matrix element $|V_{ud}|$ and of the transverse and longitudinal spectral form factors. We are now going to discuss how Eq.\,\eqref{Rtau} can be evaluated on the lattice, using the HLT method of Ref.~\cite{Hansen:2019idp}.

\subsection{The smeared $R_{ud}^{(\tau)}(\sigma)$ via the HLT method}
In this section we illustrate how the HLT method of Ref.~\cite{Hansen:2019idp} can be applied to evaluate the inclusive hadronic $\tau$-decay expressed by the ratio $R_{ud}^{(\tau)}$ of Eq.~(\ref{Rtaudef}). To this end, let us start from Eq.~(\ref{Rtau}), which can equivalently be written as
\begin{align}
\label{eq:RandKernel}
R_{ud}^{(\tau)} = 12 \pi \, S_{EW} \, \frac{|V_{ud}|^2}{m_{\tau}^{3}}  \int_{0}^{\infty} dE \left[ K_\mathrm{T}\left(\frac{E}{m_{\tau}}\right)E^2\rho_{\mathrm{T}}(E^{2}) + K_\mathrm{L}\left(\frac{E}{m_{\tau}}\right)E^2\rho_{\mathrm{L}}(E^{2})\right]~,
\end{align}
where $E = m_{\tau}\sqrt{s}$ and we have introduced the kernel functions
\begin{align}
\label{eq:kernel}
K_\mathrm{L}(x) \equiv \frac{1}{x}\left( 1 - x^{2} \right)^{2}\theta( 1 -x)~, \qquad\qquad K_\mathrm{T}(x) \equiv \left(1 + 2x^{2}\right)K_{L}(x)~.  %\sqrt{s}=\frac{E}{m_\tau}\equiv x~.
\end{align}
As it will be described in more details in the next section, the quantity we directly compute through numerical simulations is the Euclidean correlator $C^{\alpha\beta}(t,\bs{0})$ at vanishing spatial momentum $\bs{q}=\bs{0}$. For $q^{\mu} = (q, \bs{0})$, using the relations
\begin{align}
\rho^{00}(q) = q^{2}\rho_{\mathrm{L}}(q^{2})~,\qquad\qquad \frac{1}{3}\sum_{j=1,2,3}\rho^{jj}(q) = q^{2}\rho_{\mathrm{T}}(q^{2})~,
\end{align}
one has that the transverse and longitudinal form factors $\rho_{\mathrm{T}}(E^{2})$ and $\rho_{\mathrm{L}}(E^{2})$ are related to $C^{\alpha\beta}(t, \bs{0})$ through (see also Eq.~(\ref{CtE}))
\begin{align}
\label{eq:Lapl_inv}
C_\mathrm{I}(t) \equiv \int_{0}^{\infty} \frac{ dE}{2\pi}\, e^{-Et}\, E^2 \rho_\mathrm{I}(E^{2})~,
\end{align}
where we introduced the index $\mathrm{I}=\{\mathrm{L},\mathrm{T}\}$, 
and have defined 
\begin{align}
C_\mathrm{L}(t) \equiv C^{00}(t,\bs{0})\; ,
\qquad\qquad
C_\mathrm{T}(t) \equiv \frac{1}{3}\sum_{j=1,2,3} C^{jj}(t,\bs{0})\; .
\end{align}

As discussed in detail in Refs.~\cite{Hansen:2019idp,Bulava:2021fre}, the problem of reconstructing the convolution integral between spectral densities and smooth analytic kernel functions from the knowledge of the corresponding lattice Euclidean correlators, is well-posed, and it can be shown~\cite{Bulava:2021fre} that the finite-size effects (FSEs) on the resulting convolution integral vanish faster than any power of the lattice spatial extent $L$.  
In our case, the kernel functions $K_\mathrm{I}(x)$, due to the presence of the theta function, have a non-analiticity at $x=1$, and the expression in Eq.~(\ref{eq:RandKernel}) needs to be regularized. In this work we consider the following smeared kernel functions
\begin{align}
\label{eq:sm_kernels}
K_\mathrm{L}^{\sigma}(x) \equiv \frac{1}{x}\left( 1 - x^{2} \right)^{2}\Theta_{\sigma}( 1 -x)~, \qquad\qquad K_\mathrm{T}^{\sigma}(x) \equiv \left(1 + 2x^{2}\right)K_\mathrm{L}^{\sigma}(x)~,
\end{align}
where 
\begin{align}
\label{eq:sigmoid}
\Theta_{\sigma}(x) \equiv \frac{1}{1+e^{-x/\sigma}}\;,
\qquad
\qquad
\lim_{\sigma\mapsto 0}\Theta_\sigma(x)=\theta(x)\; ,
\end{align}
is a $C^{\infty}$ smoothed version of the Heaviside step-function $\theta(x)$. Clearly one has
\begin{align}
\label{eq:R_sigma}
R_{ud}^{(\tau)}  = \lim_{\sigma\mapsto 0} R_{ud}^{(\tau)}(\sigma), \qquad R_{ud}^{(\tau)}(\sigma) \equiv  12 \pi \, S_{EW} \, |V_{ud}|^2  \int_{0}^{\infty} \frac{dE\, E^2}{m_{\tau}^{3}} \left[ K_\mathrm{T}^{\sigma}\left(\frac{E}{m_{\tau}}\right)\rho_{\mathrm{T}}(E^{2}) + K_\mathrm{L}^{\sigma}\left(\frac{E}{m_{\tau}}\right)\rho_{\mathrm{L}}(E^{2})\right]~,
\end{align}
and, by performing the $L\mapsto\infty$ extrapolation at fixed non-zero values of $\sigma$ (i.e. by taking the $L\mapsto\infty$ limit \textit{before} $\sigma\mapsto 0$), we are guaranteed to deal with quantities that have a well defined infinite-volume limit and FSEs vanishing faster than any power of $1/L$.

The representation of $R_{ud}^{(\tau)}(\sigma)$ given in Eq.~(\ref{eq:R_sigma}) allows for a straightforward application of the HLT reconstruction method developed in Ref.~\cite{Hansen:2019idp}, and recently applied to the study of many hadronic processes~\cite{Gambino:2020crt, Alexandrou:2022tyn, Frezzotti:2023nun,Bonanno:2023ljc,Barone:2023tbl}. We summarize here the main ingredients of the procedure. The final goal is to find, for fixed non-zero values of the smearing parameter $\sigma$, the best approximation of the kernel functions $K_\mathrm{I}^{\sigma}$, in terms of the basis function\footnote{On a lattice having a finite temporal extent $T$, the basis function gets modified replacing $e^{-a E n}$ with $e^{-a E n} + e^{-a E (T/a-n)}$.} $\{ e^{-aE n}
\}_{n=1,\ldots, n_{max}}$, namely
\begin{align}
\label{eq:kernel_expansion}
K^{\sigma}_\mathrm{I}\left(\frac{E}{m_{\tau}}\right)  \simeq   \sum_{n=1}^{n_{max}} g_\mathrm{I}(n,\sigma, am_{\tau})\, e^{-a E n} \equiv \widetilde{K}^{\sigma}_\mathrm{I}\left(\frac{E}{m_{\tau}}\right)~,
\end{align}
where the dimension $n_{max}$ of the exponential basis is typically chosen to be equal to the number of discrete times at which $C_\mathrm{I}(t)$ is known, i.e. $n_{max}= T/(2a)$ (due to the invariance of the correlator under $t\to T-t$). In this way, once the coefficients $g_\mathrm{I}$ are known, the smeared ratio $R_{ud}^{(\tau)}(\sigma)$ can be reconstructed, from the knowledge of $C_\mathrm{I}(na)$ only, using
\begin{align}
\label{eq:HLT_master}
R_{ud}^{(\tau, \mathrm{I})}(\sigma) &\equiv 12 \pi \, S_{EW} \, |V_{ud}|^2  \int_{m_{h}}^{\infty} \frac{dE\, E^2}{m_{\tau}^{3}} \,K_\mathrm{I}^{\sigma}\left(\frac{E}{m_{\tau}}\right)\rho_\mathrm{I}(E^{2})
\nonumber \\[8pt]
&\simeq 12 \pi \, S_{EW} \, |V_{ud}|^2 \sum_{n=1}^{T/(2a)} g_{I}(n, \sigma, am_{\tau}) \int_{0}^{\infty} \frac{dE\, E^2}{m_{\tau}^{3}}\,  e^{-a E n }\rho_\mathrm{I}(E^{2}) \nonumber \\[8pt]
&= \frac {24 \pi^{2}}{m_{\tau}^{3}} \, S_{EW} \, |V_{ud}|^2 \sum_{n=1}^{T/(2a)} g_\mathrm{I}(n,\sigma, am_{\tau})\, C_\mathrm{I}(na)~,
\end{align}
and clearly $R_{ud}^{(\tau)}(\sigma) = R_{ud}^{(\tau, \mathrm{T})}(\sigma) + R_{ud}^{(\tau, \mathrm{L})}(\sigma)$. 

The problem of finding the coefficients $g_\mathrm{I}$ presents a certain number of technical difficulties. Any determination of the smeared ratio $R_{ud}^{(\tau)}(\sigma)$ based on Eqs.~(\ref{eq:kernel_expansion}),~(\ref{eq:HLT_master}) will be inevitably affected by both systematic errors (due to the inexact reconstruction of the kernels) and statistical uncertainties (due to the fluctuations of the correlator $C_\mathrm{I}(t)$), which need to be simultaneously kept under control. 
The HLT method enables to find an optimal balance between the size of the statistical and systematic errors. This is achieved by minimizing a linear combination
\begin{align}
\label{eq:func_W}
W_\mathrm{I}^{\alpha}[\boldsymbol{g}] \equiv \frac{A_\mathrm{I}^{\alpha}[\boldsymbol{g}]}{A_\mathrm{I}^{\alpha}[\boldsymbol{0}]} + \lambda B_\mathrm{I}[\boldsymbol{g}]\;,
\end{align}
of the norm-functional
\begin{align}
\label{eq:func_A}
A_\mathrm{I}^{\alpha}[\bs{g}] = \int_{E_{min}}^{E_{max}} dE\, e^{a E \alpha}\, \bigg| \sum_{n=1}^{T/(2a)} g(n) e^{-a E n} - K_\mathrm{I}^{\sigma}\left(\frac{E}{m_{\tau}}\right) \bigg |^{2}~,\qquad E_{max}\equiv r_{max}/a~, 
\end{align}
which quantifies the difference between the approximated and the target kernel, 
and of the error-functional
\begin{align}
\label{eq:func_B}
B_\mathrm{I}[\boldsymbol{g}] =  B_{\rm{norm}}\sum_{n_1 , n_2 = 1}^{T/(2a)} g(n_{1}) ~ g(n_{2})~ {\rm{Cov}_\mathrm{I}}(an_1 , an_2 )~, 
\end{align}
where ${\rm Cov}_\mathrm{I}$ is the covariance matix of the correlator $C_\mathrm{I}(an)$, and $B_{\rm norm}$ is a normalization factor introduced to render the error-functional dimensionless. The algorithmic parameters $\{ \alpha, E_{min}, r_{max} \}$ in Eq.~(\ref{eq:func_A}) can be tuned to optimize the fidelity of the reconstruction in specific energy regions. For the parameter $E_{min}$, any value $0 < E_{min} < m_{h}$, where $m_{h}$ is 
the mass of the lightest hadronic state allowed in the decay (in pure QCD $m_h = m_\pi$ for the axial channel and $m_h = 2\, m_\pi$ for the vector one), is legitimate, since the spectral density vanishes for $E \in [0, m_{h})$, and the quality of the reconstruction of the kernel functions in this energy region has no impact in the determination of $R_{ud}^{(\tau,\mathrm{I})}(\sigma)$.
The parameter $\lambda$ in Eq.~(\ref{eq:func_W}) is the so-called \textit{trade-off} parameter, and for a given value of $\lambda$, the minimization of the functional $W_\mathrm{I}^{\alpha}[\bs{g}]$ gives the coefficients $\bs{g^{\lambda}_\mathrm{I}}$. In presence of statistical errors, the second term in Eq.~(\ref{eq:func_W}) disfavours coefficients $\bs{g}$ leading to too large statistical uncertainties in the reconstructed values of $R_{ud}^{(\tau,\mathrm{I})}$. The optimal balance between having small statistical errors (small $B_\mathrm{I}[\bs{g}]$) and small systematic errors (small $A_\mathrm{I}[\bs{g}]$) can be achieved by tuning $\lambda$ appropriately. This is done performing the so-called \textit{stability-analysis}, which is thoroughly discussed in Refs.~\cite{Alexandrou:2022tyn, Frezzotti:2023nun}. In a nutshell, in the stability-analysis one monitors the evolution of the reconstructed values of $R_{ud}^{(\tau,\mathrm{I})}$ as a function of $\lambda$. The optimal value $\lambda^{\filledstar}$ is then chosen in the so-called statistically-dominated regime, where $\lambda$ is sufficiently small that the systematic error due to the kernel reconstruction is smaller than the statistical one (in this region the results are thus stable under variations of $\lambda$), but large enough to still have reasonable statistical uncertainties. Examples of the stability analysis will be given in the next section (see Fig.~\ref{fig:stab_plot}). Finally, having determined the optimal value $\lambda^{\filledstar}$, we always repeat the calculation employing a second (smaller) value of $\lambda = \lambda^{\filledstar\filledstar}$, which is determined imposing the validity of the following condition
\begin{align}
\label{eq:lambda_syst}
\frac{B_\mathrm{I}[\bs{g_\mathrm{I}^{\lambda^{\filledstar\filledstar}}}]}{A_\mathrm{I}[ \bs{g^{\lambda^{\filledstar\filledstar}}_\mathrm{I}}]} = \kappa\, \frac{B_\mathrm{I}[\bs{g_\mathrm{I}^{\lambda^{\filledstar}}}]}{A_\mathrm{I}[ \bs{g_\mathrm{I}^{\lambda^{\filledstar}}}]}~,
\end{align}
with $\kappa=10$. Any statistically-significant difference between the values of $R_{ud}^{(\tau,\mathrm{I})}(\sigma)$ corresponding to the two choices $\lambda=\lambda^{\filledstar}$ and $\lambda=\lambda^{\filledstar\filledstar}$, is added as a systematic uncertainty in our final error. We refer to Ref.~\cite{Alexandrou:2022tyn} for further details on this point.

\section{Numerical results}
\label{sec:numerical_results}

In this section we show our numerical results for $R_{ud}^{(\tau)}$, obtained making use of the last generation of gauge field configurations produced by the Extended Twisted Mass Collaboration (ETMC) employing the Iwasaki gluon action~\cite{Iwasaki:1985we} and $N_{f}=2+1+1$ flavours of Wilson-Clover twisted-mass fermions at maximal twist~\cite{Frezzotti:2000nk}. This framework guarantees the automatic $\mathcal{O}(a)$ improvement of 
all physical observables of interest~\cite{Frezzotti:2003ni,Frezzotti:2004wz}.
Full information on the last generation of ETMC ensembles can be found in Refs.~\cite{ExtendedTwistedMass:2021gbo,ExtendedTwistedMass:2021qui,ExtendedTwistedMass:2022jpw,Alexandrou:2018egz}, as well as in Refs.~\cite{Finkenrath:2022eon, Kostrzewa:2022hsv} for more technical details. 
\begin{table}
\begin{center}
    \begin{tabular}{||c||c|c|c|c|c||c||}
    \hline
    ~~~ ensemble ~~~ & ~~~ $\beta$ ~~~ & ~~~ $V/a^{4}$ ~~~ & ~~~ $a$\,(fm) ~~~  & ~~~ $M_{\pi}$\,(MeV) ~~~ & ~~~ $L$ (fm) ~~~ & ~~~ $N_{g}$ ~~~\\
  \hline
  
  B64 & $1.778$ & $64^{3}\cdot 128$ & $0.07957~(13)$ &  $135.2~(0.2)$ & $5.09$ & $776$ \\

 B96 & - & $96^{3}\cdot 192$ & - & - & $7.64$ & $602$ \\ 
  
  C80 & $1.836$ & $80^{3}\cdot 160$ & $0.06821~(13)$ &  $134.9~(0.3)$ & $5.46$ & $401$  \\
  
  D96 & $1.900$ & $96^{3}\cdot 192$ & $0.05692~(12)$ &  $135.1~(0.3)$ & $5.46$ & $373$ \\
  \hline
    \end{tabular}
\end{center}
\caption{\it \small Parameters of the ETMC ensembles used in this work. We present  the lattice spacing $a$,  the pion mass $M_\pi$, the lattice size $L$, and the number $N_{g}$ of gauge configurations analyzed. The values of the lattice spacing are determined as explained in Appendix B of Ref.~\cite{ExtendedTwistedMass:2022jpw} using the value $f_\pi^{phys} = f_\pi^{isoQCD} = 130.4(2)~{\rm MeV}$ of the pion decay constant. The quoted values of the pion mass are obtained by computing the light-quark mass corrections needed to match the value $M_{\pi}^{isoQCD} = 135.0~{\rm MeV}$ starting from simulations with slightly larger than physical quark masses, as described in the Appendix A of Ref.~\cite{ExtendedTwistedMass:2022jpw}. We refer to Ref.~\cite{ExtendedTwistedMass:2022jpw} for more detailed information on the ensembles.}
\label{tab:simudetails}
\end{table} 
However, for the sake of completeness, we report in Tab.~\ref{tab:simudetails} essential information on the ensembles that have been used for the present work, which correspond to three values of the lattice spacing $a$ in the range $[0.056,0.08]\,{\rm fm}$, and lattice extent $L$ in the range $[5.09, 7.64]\,{\rm fm}$. All ensembles but the B96 have a very similar volume (see Tab.~\ref{tab:simudetails}), while the B96 ensemble has a larger linear extent of more than $7~{\rm fm}$, and it is used to estimate finite-size effects (FSEs).

We employ two distinct lattice discretizations of the weak current, given by
\begin{align}
\label{eq:weak_current_latt}
J_{ud}^{\alpha, {\rm tm}}= Z_{A}\bar{u}_{+}\gamma^{\alpha}d_{+}  - Z_{V}\bar{u}_{+}\gamma^{\alpha}\gamma_{5}d_{+}~, \qquad\qquad J_{ud}^{\alpha, {\rm OS}} = Z_{V}\bar{u}_{+}\gamma^{\alpha}d_{-} - Z_{A}\bar{u}_{+}\gamma^{\alpha}\gamma_{5}d_{-}~,
\end{align}
where $Z_{A}$ and $Z_{V}$ are the (finite) renormalization constants (RCs) of the axial and vector currents, which in twisted-mass QCD are chirally rotated with respect to the ones of standard Wilson fermions. In the previous equation, the subscript $\pm$ on the up and down quark-fields, corresponds to a specific choice of the Wilson $r-$parameter (see e.g.~\cite{Alexandrou:2022amy} and reference therein for details), given by 
\begin{align}
r_{u_{\pm}} = -r_{d_{\pm}} = (-1)^{\pm}~.
\end{align}
Since the twisted Wilson term accompanied by the appropriate critical mass counterterm (for both sea
and all kind of valence quark fields) is a truly dimension-five irrelevant operator, the two currents $J_{ud}^{{\rm tm}}$ and $J_{ud}^{{\rm OS}}$, where ``tm'' stands for twisted-mass and ``OS'' for Osterwalder-Seiler, when plugged into the expression~(\ref{Ctt}), produce Euclidean correlators having the same continuum limit. 
However, at finite lattice spacing the two correlators differ in general by lattice artifacts, allowing us to approach the continuum limit from two different directions.
In the following, we will present the results obtained using both the ``tm'' and the ``OS'' current, which, as it will be discussed, can simultaneously be used in the continuum fits to better control the result of the extrapolation. Moreover, we will oftentimes adopt the notation $X^{\rm tm\, (OS)}$, to indicate that a given lattice observable $X$ has been evaluated using the current $J_{ud}^{\rm tm\,(OS)}$. For completeness, in Tab.~\ref{tab:renormalization} we collect the values of $Z_{A}$ and $Z_{V}$ that we use  for each of the ensembles of Tab.~\ref{tab:simudetails}.
\begin{table}[]
    \centering
    \begin{tabular}{||c||c|c||}
    \hline
    ensemble & $Z_{V}$ & $Z_{A}$ \\
    \hline
    ~ B64 ~ & ~ $0.706379~(24)$ ~ & ~ $0.74294~(24)$ ~   \\
    
    ~ B96 ~ & ~ $0.706405~(17)$ ~ & ~ $0.74267~(17)$ ~  \\
  
    ~ C80 ~ & ~ $0.725404~(19)$ ~ & ~ $0.75830~(16)$ ~ \\
    
    ~ D96 ~ & ~ $0.744108~(12)$ ~ & ~ $0.77395~(12)$ ~  \\
    \hline
    \end{tabular}
    \caption{\it \small The values of the vector ($Z_{V}$) and axial ($Z_{A}$) RCs for the ETMC ensembles of Tab.\,\ref{tab:simudetails}. They have been determined in Ref.~\cite{ExtendedTwistedMass:2022jpw}, to which we refer for additional details, using 
    methods based on lattice Ward identities (for $Z_V$) or universality of matrix elements (case of $Z_A$) for currents between the vacuum and a (relatively light) one-pseudoscalar-meson state.}   
%    Ward-identity-based methods. }  % RF: il metodo per Z_A non e` WI-based...
    \label{tab:renormalization} 
\end{table}

For both the ``tm'' and the ``OS'' regularization, and for each of the ensembles of Tab.~\ref{tab:simudetails}, we evaluated the Euclidean correlator $C^{\alpha\beta}(t,\bs{0})$ at vanishing spatial momentum $\bs{q}=0$, using $N_{source}=10^{3}$ stochastic spatial sources per gauge configuration. The sources are randomly distributed in the time-slice, diagonal in spin and diluted in the color variable. The number of gauge configurations we analyzed is given in Tab.~\ref{tab:simudetails}. From the knowledge of $C^{\alpha\beta}(t,\bs{0})$, we determine the smeared ratio $R_{ud}^{(\tau)}(\sigma)$ using the HLT method introduced in the previous section, as we are now going to illustrate.

\subsection{Stability analysis and study of the FSEs}
For each of the ensembles of Tab.~\ref{tab:simudetails}, and for both the ``tm'' and the ``OS'' regularization, we evaluated $R_{ud}^{(\tau, \mathrm{I})}(\sigma)$ for several values of $\sigma \in [0.004, 0.14]$. A first investigation we carried out concerns the determination of the optimal values of the algorithmic parameters $\{\alpha, E_{min}, r_{max}\}$ of Eq.~(\ref{eq:func_A}). For $E_{min}$, we always set its value to $E_{min} = 0.05\, m_{\tau}\simeq 90~{\rm MeV}$, which ensures the validity of the condition $E_{min} < m_{h}$ (see text below Eq.~(\ref{eq:func_B})) in all channels. We considered instead different values of $\alpha$ and $r_{max}$ given by\footnote{$\alpha=2^{-}$ in practice means $\alpha=1.99$.} $\alpha \in \{ 2^{-}, 3, 4, 5\}$ and $r_{max} \in \{4, 5, 6, \infty\}$. For $E_{max} = \infty$ only values of $\alpha < 2$ can be employed, otherwise the integral in Eq.~(\ref{eq:func_A}) would be divergent.

\begin{figure}
    \centering
    \includegraphics[scale=0.45]{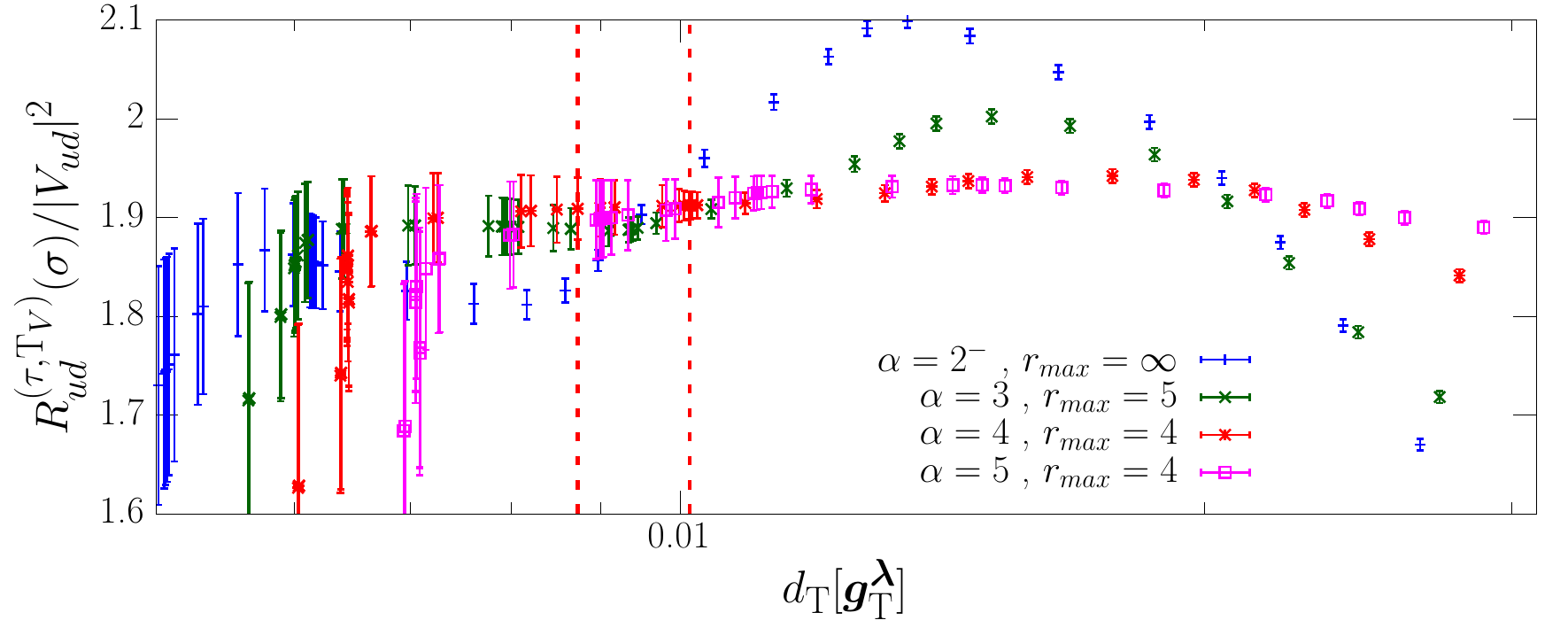}
    \includegraphics[scale=0.45]{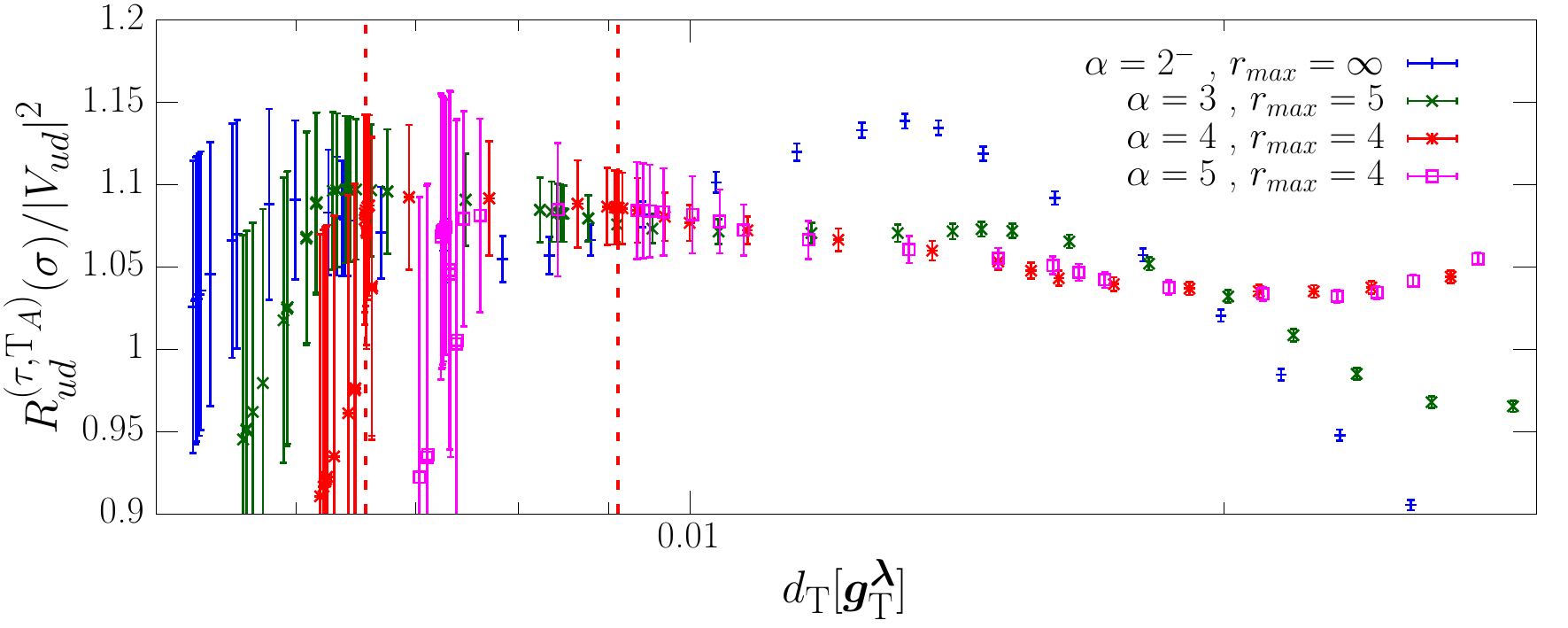}
    \includegraphics[scale=0.45]{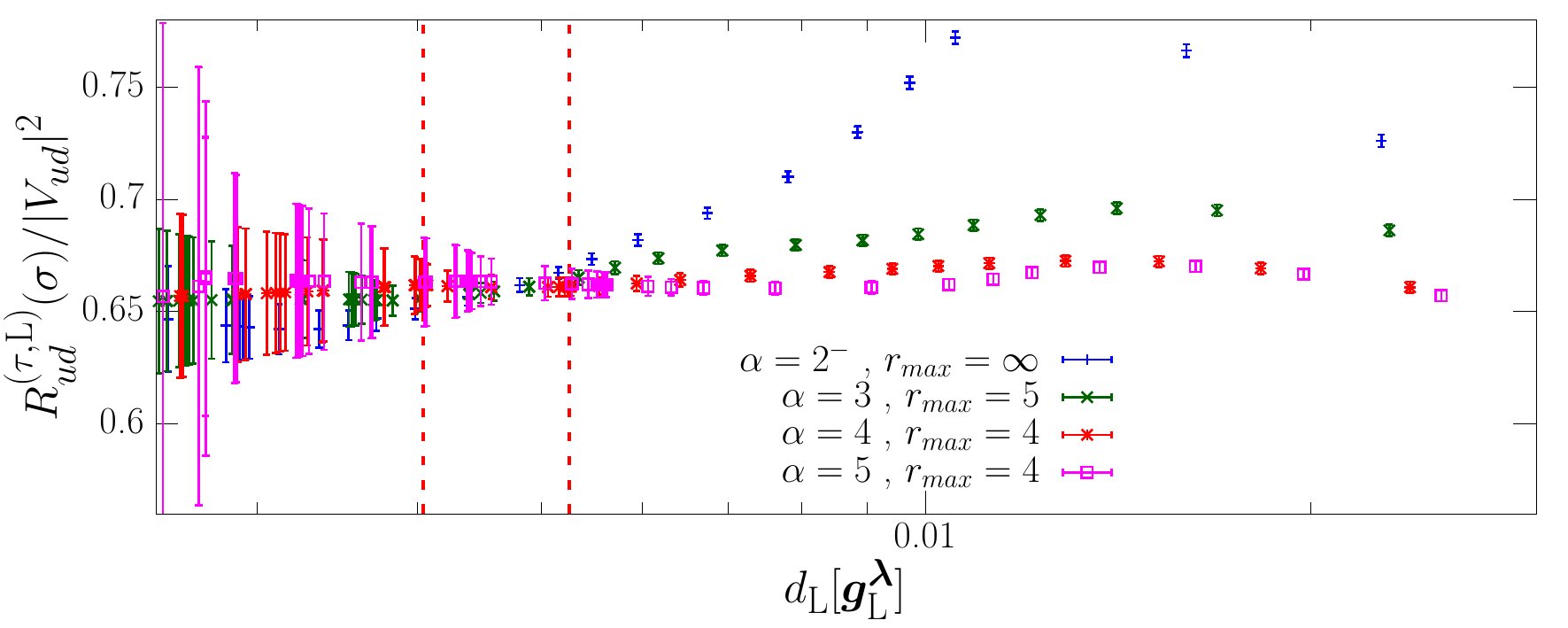}
    \caption{\small\it Stability-analysis plot for the three different contributions $R_{ud}^{(\tau, \mathrm{T}_{V})}(\sigma)$ (top figure), $R_{ud}^{(\tau, \mathrm{T}_{A})}(\sigma)$ (middle figure), and $R_{ud}^{(\tau, \mathrm{L})}(\sigma)$ (bottom figure). The data refer to the results obtained on the B64 ensemble for $\sigma= 0.02$, using the ``OS'' regularization. In each figure, the different colors correspond to different values of the algorithmic parameters $\alpha$ and $r_{max}$. The vertical dashed lines correspond, from the right to the left, to the position of the values of $\lambda^{\filledstar}$ and $\lambda^{\filledstar\filledstar}$ (see Eq.~(\ref{eq:lambda_syst}) and the text around it)  for the case $\alpha=4$ and $r_{max}=4$ (data points in red).
    }
    \label{fig:stab_plot}
\end{figure}

In Fig.~\ref{fig:stab_plot}, we show the stability-analysis plot for a fixed value of $\sigma= 0.02$ and for a selected gauge ensemble, the B64. In the figure we compare the results obtained for different choices of $\alpha$ and $E_{max}$.
Notice that we distinguish the axial and vector contributions to $R_{ud}^{(\tau,\mathrm{T})}(\sigma)$, which we indicate simply by $R_{ud}^{(\tau, \mathrm{T}_{V})}(\sigma)$ and $R_{ud}^{(\tau, \mathrm{T}_{A})}(\sigma)$. The two contributions are obtained by evaluating the Euclidean correlator of Eq.~(\ref{Ctt}) employing respectively the axial and vector part of the weak current of Eqs.~(\ref{weak_current}) and~(\ref{eq:weak_current_latt}), and then following the procedure described in the previous subsections to obtain the smeared ratio. One has
\begin{align}
  R_{ud}^{(\tau, \mathrm{T})}(\sigma) = R_{ud}^{(\tau, \mathrm{T}_{V})}(\sigma) + R_{ud}^{(\tau, \mathrm{T}_{A})}(\sigma)~,  
\end{align}
because the mixed axial-vector term in Eq.~(\ref{Cdit}) is zero due to parity symmetry. Instead, in isosymmetric QCD, the vector part of the correlation function of Eq.~(\ref{Cdit}) is transverse due to the conservation of the vector current $\bar{u}\gamma^{\alpha}d$, therefore the longitudinal contribution $R_{ud}^{(\tau, \mathrm{L})}(\sigma)$ comes entirely from the axial part of the weak current. Computing the axial and vector contributions separately is phenomenologically interesting, since, as we shall discuss, the same separation can 
be performed experimentally. 

\begin{figure}
    \centering
    \includegraphics[scale=0.43]{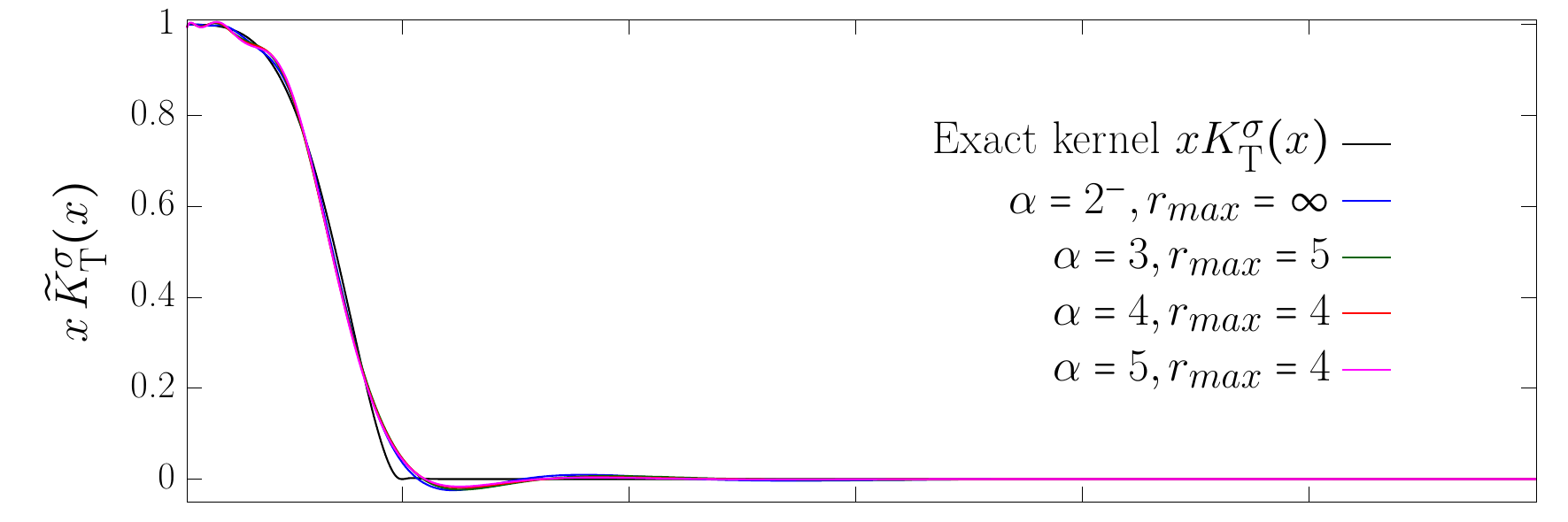}\\
    \vspace{-0.3cm}
    \includegraphics[scale=0.43]{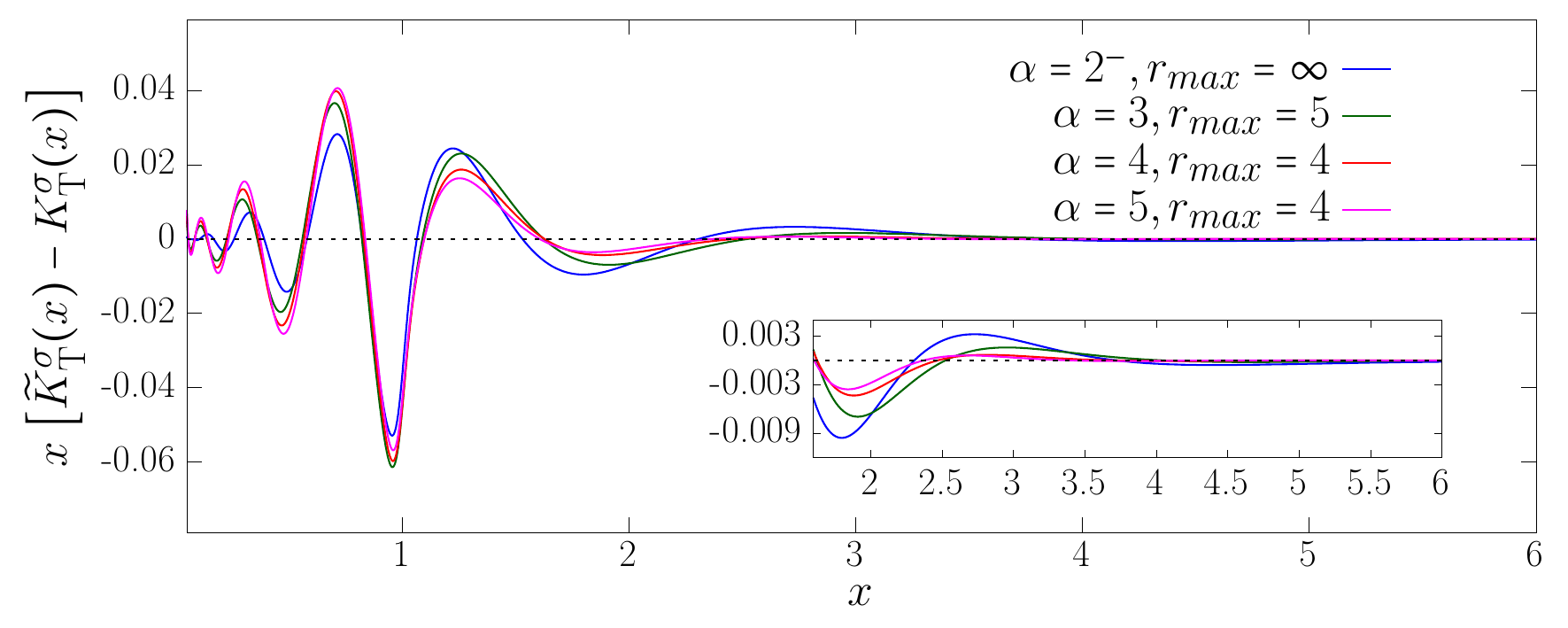}
    \caption{\small\it The reconstructed smearing kernels $\widetilde{K}_\mathrm{T}^{\sigma}(x)$ obtained using the coefficients $\bs{g_\mathrm{T}^{\lambda^{\filledstar}}}$ employed in the reconstruction of $R_{ud}^{(\tau, \mathrm{T}_{A})}(\sigma)$ in the ``OS'' regularization. The different colors correspond to the different choices of the algorithmic parameters $\alpha$ and $r_{max}$ shown in Fig.~\ref{fig:stab_plot}. In the top figure the black line corresponds to the exact kernel $K_\mathrm{T}^{\sigma}(x)$ of Eq.~(\ref{eq:sm_kernels}). In the bottom figure we show instead $x \left [ \widetilde{K}_\mathrm{T}^{\sigma}(x) - K_\mathrm{T}^{\sigma}(x)  \right ]$. The data correspond to $\sigma=0.02$. }
    \label{fig:kernel_reco}
\end{figure}

In Fig.~\ref{fig:stab_plot} the results corresponding to the different values of $\lambda$ explored are shown as a function of the variable
\begin{align}
d_\mathrm{I}[\bs{g_\mathrm{I}^{\lambda}}] \equiv \sqrt{ \frac{A^{0}_\mathrm{I}[\bs{g_\mathrm{I}^{\lambda}}]}{A^{0}_\mathrm{I}[\bs{0}]} }
\end{align}
which quantifies the goodness of the reconstruction (the smaller the value of $d_\mathrm{I}[\bs{g_\mathrm{I}^{\lambda}}]$, the better the reconstruction). As the figure shows, larger values of $\alpha$ lead to much more stable results as a function of $\lambda$. This behaviour is explained by noticing that for $\alpha>0$ the presence of the exponential $e^{a E \alpha}$ term in Eq.~(\ref{eq:func_A}) improves the quality of the reconstruction in the large-energy region. Indeed,  the errors in the reconstruction of the smearing kernels for large values of $E/m_\tau$ get amplified in the corresponding smeared quantity because, in general, the spectral densities appearing in lattice correlators grow asymptotically with the energy. In light of these findings, in the following we employ in the data analysis only the results corresponding to $\alpha > 2$ (and hence to a finite $r_{max}$). For illustration, we report in Fig.~\ref{fig:kernel_reco} the reconstructed smeared kernels $\widetilde{K}_\mathrm{T}^{\sigma}(x)$, for $\sigma =0.02$, corresponding to  the coefficients $\bs{g_\mathrm{T}^{\lambda^{\filledstar}}}$ determined in the spectral reconstruction of $R_{ud}^{(\tau, \mathrm{T}_{A})}(\sigma)$, in the ``OS'' regularization and for all the values of the algorithmic parameters shown in Fig.~\ref{fig:stab_plot}. 

\begin{figure}
    \centering
    \includegraphics[scale=0.29]{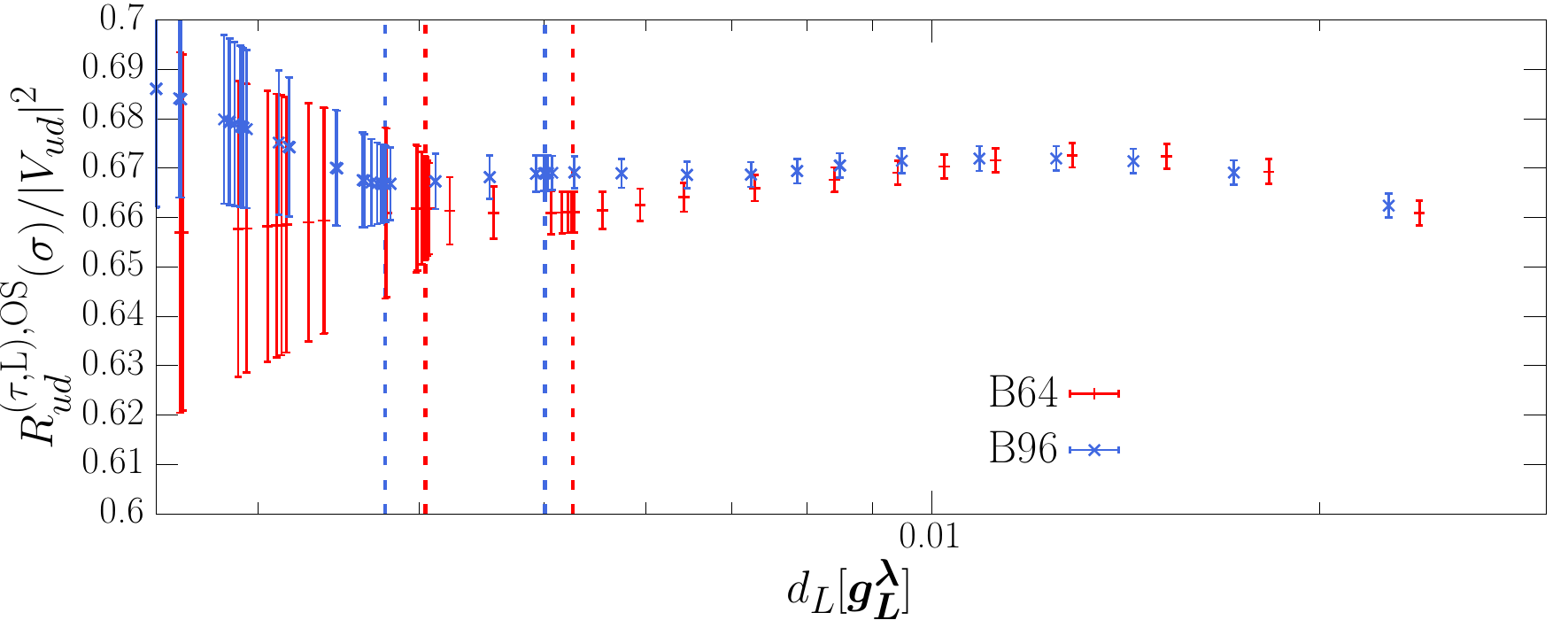}
    \includegraphics[scale=0.29]{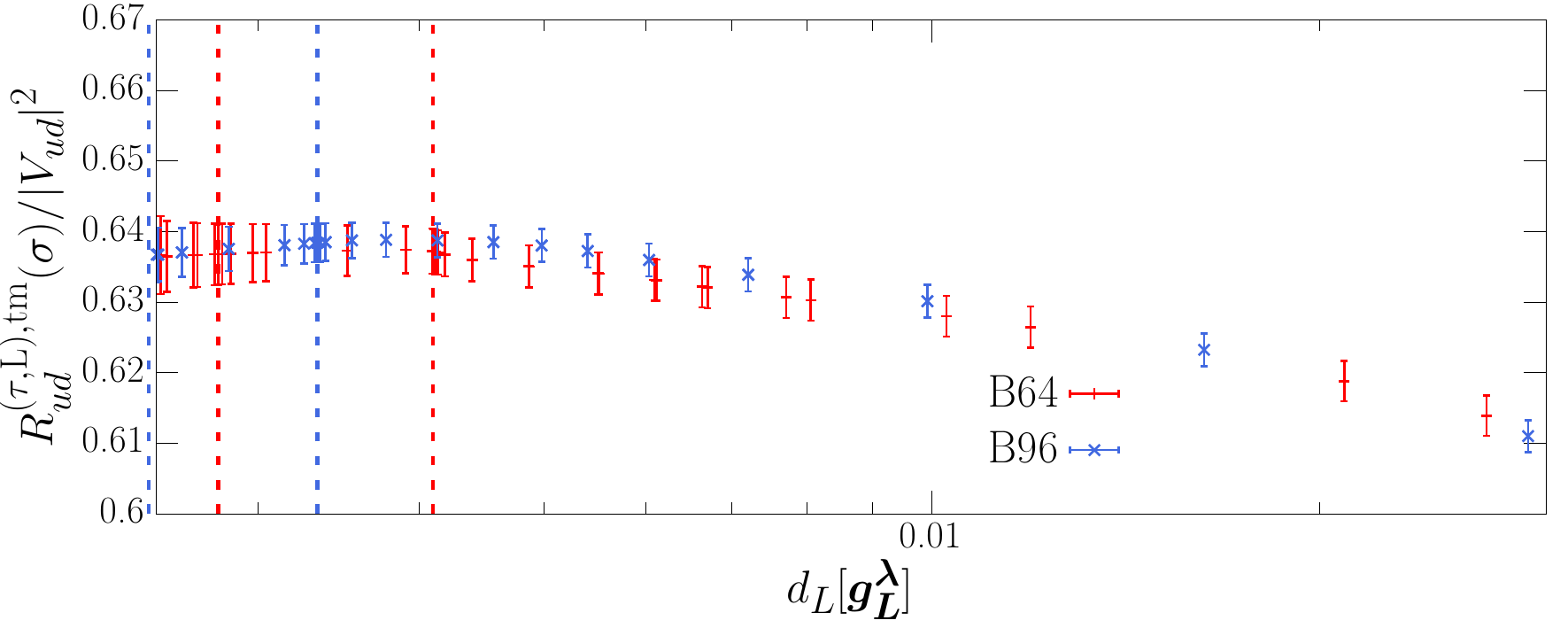}\\

    \includegraphics[scale=0.29]{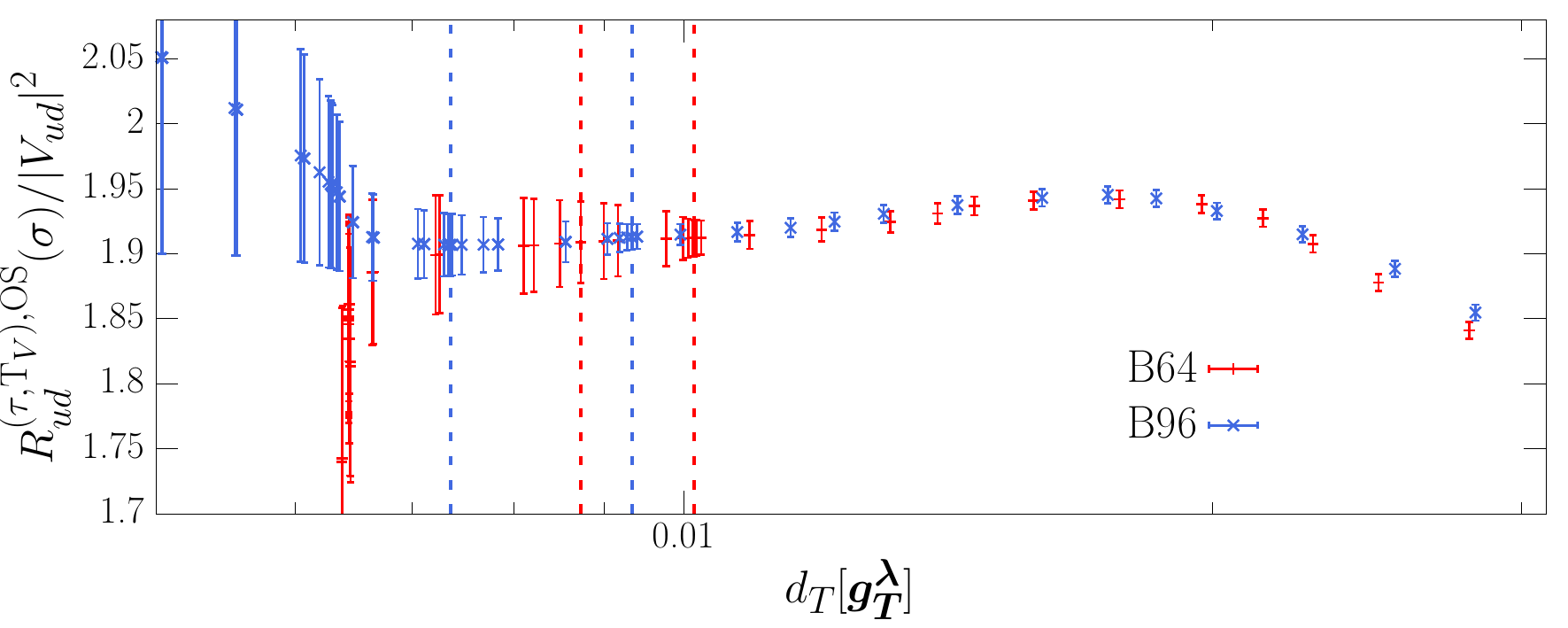}
    \includegraphics[scale=0.29]{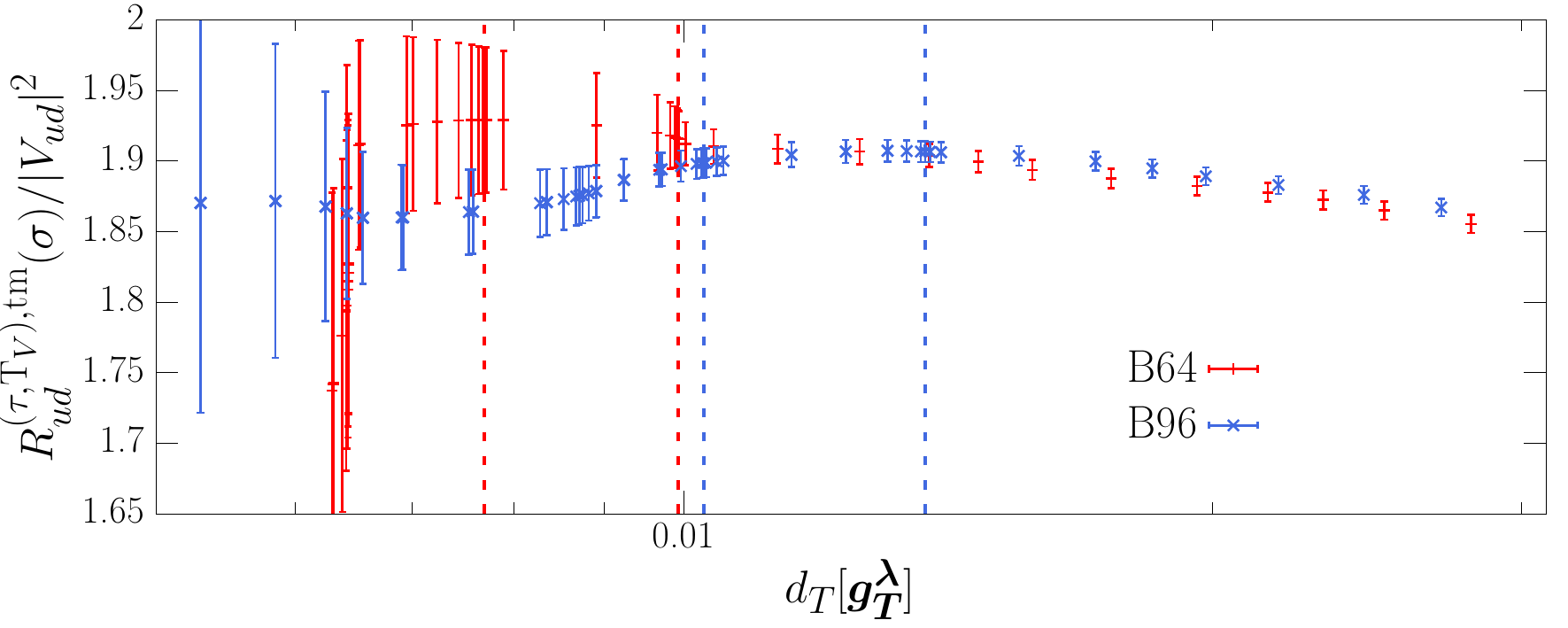}\\

    \includegraphics[scale=0.29]{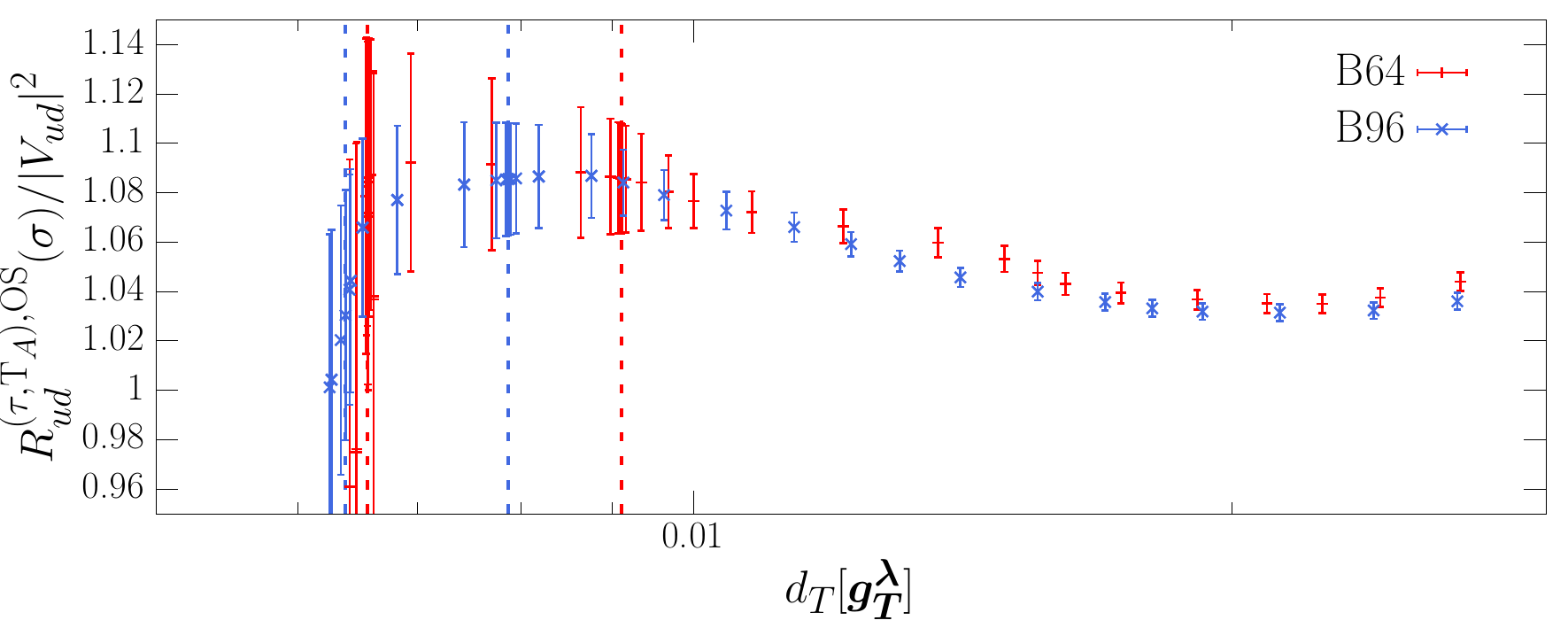}
    \includegraphics[scale=0.29]{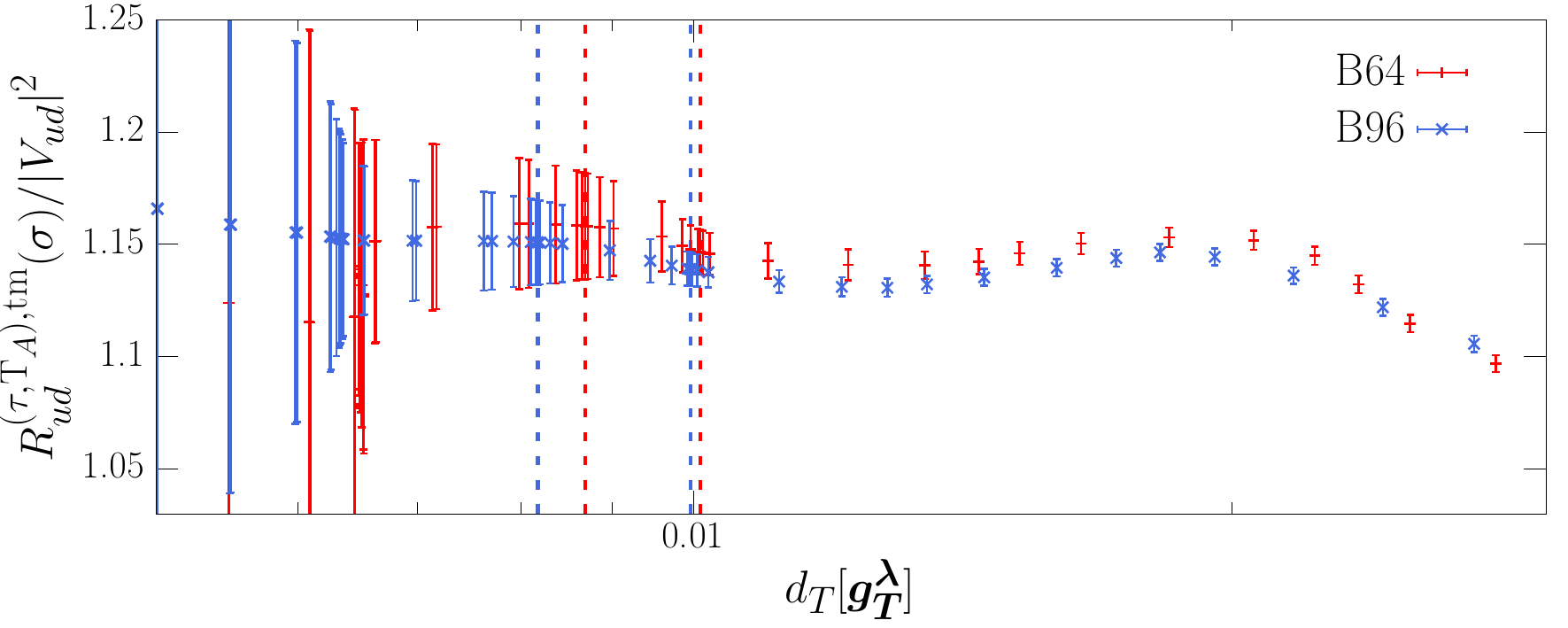}
    \caption{\small\it Comparison between the stability analysis plot corresponding to the B64 (red) and B96 (blue) ensembles. The three figures on the left (right) correspond to the different contributions to the smeared ratio for $\sigma=0.02$ and in the ``OS'' (``tm'') regularization. As in Fig.~\ref{fig:stab_plot}, the vertical dashed lines correspond, from the right to the left, to the position of the values of $\lambda^{\filledstar}$ and $\lambda^{\filledstar\filledstar}$ for the two ensembles. All data correspond to the spectral reconstruction obtained using $\alpha=4$ and $r_{max}=4$. }
    \label{fig:comp_volumes}
\end{figure}

Concerning the estimate of the FSEs, we recall that all the gauge ensembles we use but the B96 have a very similar spatial volume (in the range $L \in [5.1, 5.5]~{\rm fm}$). To properly estimate the FSEs we compare the results obtained on the B64 and B96 ensembles, which share the same values of the quark masses and of the lattice spacing, and only differ by the lattice volume. The B96 has a large lattice extent of $7.64~{\rm fm}$.  The results of the comparison are shown in Fig.~\ref{fig:comp_volumes} for $\sigma=0.02$, and for both the ``tm'' and the ``OS'' regularization, in the case of the spectral reconstruction obtained using $\alpha=4$ and $r_{max}=4$.
As the figure shows, for all contributions and regularizations, the difference between the results on the two volumes is reasurringly small. However, in order to be conservative, we associate to the results obtained on the B64, C80 and D96 ensembles an additional systematic uncertainty, due to FSEs, given by
\begin{align}
\Sigma_\mathrm{I}^{{\rm FSE}}(\sigma) = \max_{{\rm reg} = \{ {\rm tm}, {\rm OS} \}}\left\{ \Delta_\mathrm{I}^{\rm reg}(\sigma) \erf\left(\frac{1}{\sqrt{2}\sigma_{\Delta_\mathrm{I}^{\rm reg}(\sigma)}}\right) \right\}~,
\end{align}
where we have defined
\begin{align}
\Delta^{\rm reg}_\mathrm{I}(\sigma) = \bigg| R_{ud}^{(\tau,\mathrm{I}),  {\rm reg}}(\sigma, {\rm B64}) - R_{ud}^{(\tau,\mathrm{I}),  {\rm reg}}(\sigma, {\rm B96}) \bigg|~,
\end{align}
and indicated with $\sigma_{\Delta_\mathrm{I}^{\rm reg}(\sigma)}$ its relative statistical uncertainty. Thus, $\Sigma_\mathrm{I}^{\rm FSE}(\sigma)$ is the 
 spread between the results obtained on the B96 and B64 ensembles, weighted by the probability that
the spread is not due to a statistical fluctuation, and maximized over the two regularizations ``tm'' and ``OS''. In the following subsection, we will carry out the continuum limit extrapolation using the results on the B64, C80 and D96 ensembles, which after including the systematic uncertainty $\Sigma_\mathrm{I}^{\rm FSE}(\sigma)$ due to FSEs, are considered as infinite-volume quantities.

%#######################################
%#######################################
%#######################################
%#######################################
\subsection{Continuum limit extrapolation and the limit of vanishing $\sigma$}
We now turn into the discussion of the remaining extrapolations that need to be performed to obtain $R_{ud}^{(\tau)}$, namely the continuum limit extrapolation at fixed $\sigma$ (that we perform first), and the final extrapolation to vanishing $\sigma$. For the continuum extrapolation, we perform combined fits to the data corresponding to the two regularizations ``tm'' and ``OS'', employing the following fit Ansatz
\begin{align}
R_{ud}^{(\tau,\mathrm{I}), {\rm tm}}(\sigma, a) = R_\mathrm{I}(\sigma) + D_\mathrm{I}^{\rm tm}(\sigma) a^{2}~,\qquad R_{ud}^{(\tau,\mathrm{I}), {\rm OS}}(\sigma, a) = R_\mathrm{I}(\sigma) + D_\mathrm{I}^{\rm OS}(\sigma) a^{2} ~,  
\end{align}
where $R_\mathrm{I}(\sigma), D_\mathrm{I}^{\rm tm}(\sigma)$, and $D_\mathrm{I}^{\rm OS}(\sigma)$ are $\sigma-$dependent free fit parameters. A common continuum limit value, $R_\mathrm{I}(\sigma)$, is thus enforced. The fit is performed minimizing a correlated $\chi^{2}-$variable. The corresponding covariance matrix has a $2\times 2$ block-diagonal form, since the data corresponding to different ensembles are uncorrelated, and the only non-vanishing correlation is the one between the ``tm'' and ``OS'' data corresponding to the same ensemble.  We carry out, for each contribution, a total of four different fits, which differ on whether for each regularization we perform a linear or a constant fit in $a^{2}$ (i.e. we selectively set $D_\mathrm{I}^{\rm tm}(\sigma)$ and/or $D_\mathrm{I}^{\rm OS}(\sigma)$ to zero). In order to combine the results obtained in the different continuum fits, and provide our final determination of $R_{ud}^{(\tau, \mathrm{I})}(\sigma)$, we make use of the Bayesian model average (BMA) procedure developed in Ref.~\cite{EuropeanTwistedMass:2014osg}: starting from the  values $\{x_{k}\}_{k=1,\ldots, N}$ obtained by fitting the same dataset in $N$ different ways, their average $\bar{x}$ and final uncertainty $\sigma_{x}$  are given by
\begin{align}
\label{eq:BMA}
\bar{x} = \sum_{k=1}^{N} \omega_{k} x_{k}~,\qquad \sigma_{x}^{2} = \sigma_{x, stat}^{2} + \sum_{k=1}^{N} \omega_{k} (x_{k} -\bar{x})^{2}~,\qquad \sum_{k=1}^{N}\omega_{k}=1 
\end{align}
where  $\omega_{k}$ is the weight associated to the $k-$th fit, and $\sigma_{x, stat}$ is the statistical error of $\bar{x}$. We choose the weights according to the Akaike Information Criterion (AIC) of Ref.~\cite{Akaike}, namely
\begin{align}
\label{eq:AIC}
\omega_{k} \propto \exp\left\{ -\left( \chi^{2}_{k} + 2N^{k}_{par} - N^{k}_{meas}  \right)/2\right\}~,
\end{align}
where $\chi^{2}_{k}$, $N^{k}_{par}$ and $N^{k}_{meas}$ are respectively the chi-squared, the total number of fit parameters, and the total number of measurements of the $k-$th fit.

\begin{figure}
    \centering
    \includegraphics[scale=0.32]{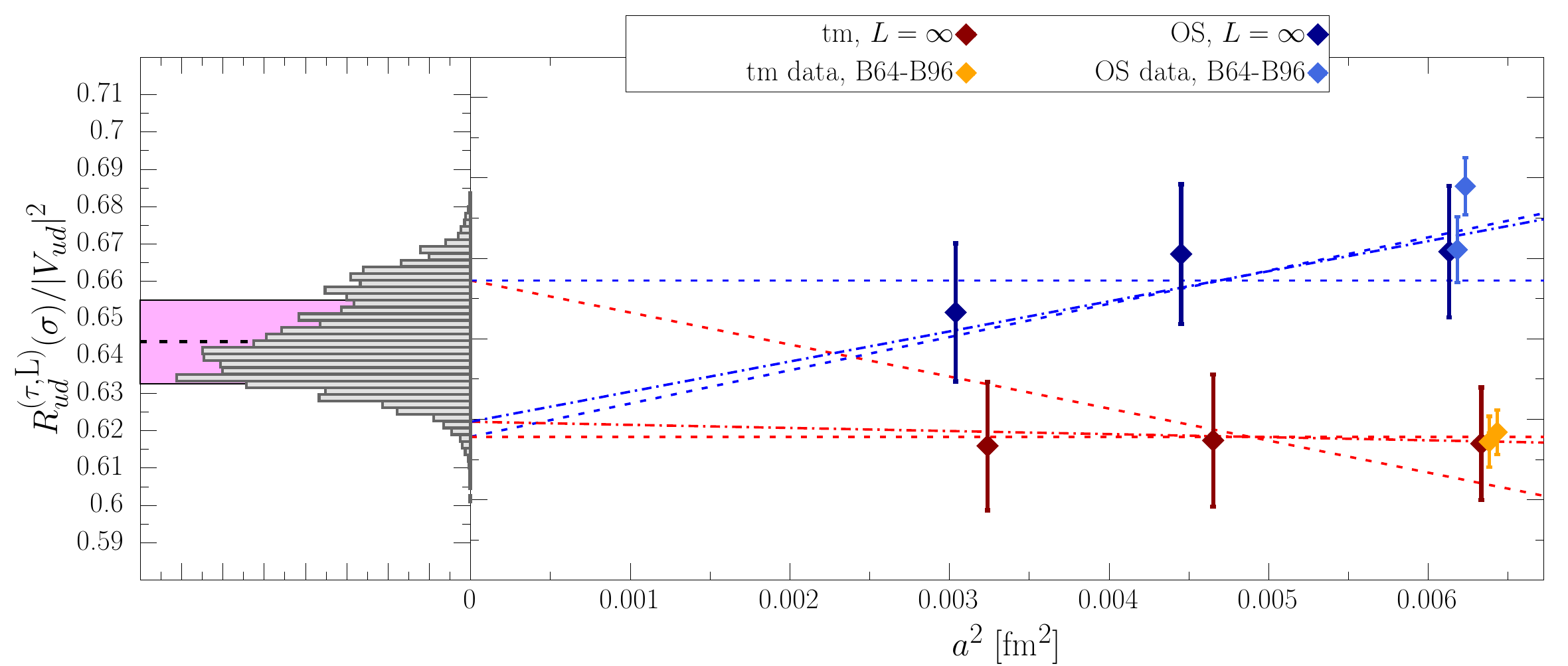}
    \includegraphics[scale=0.32]{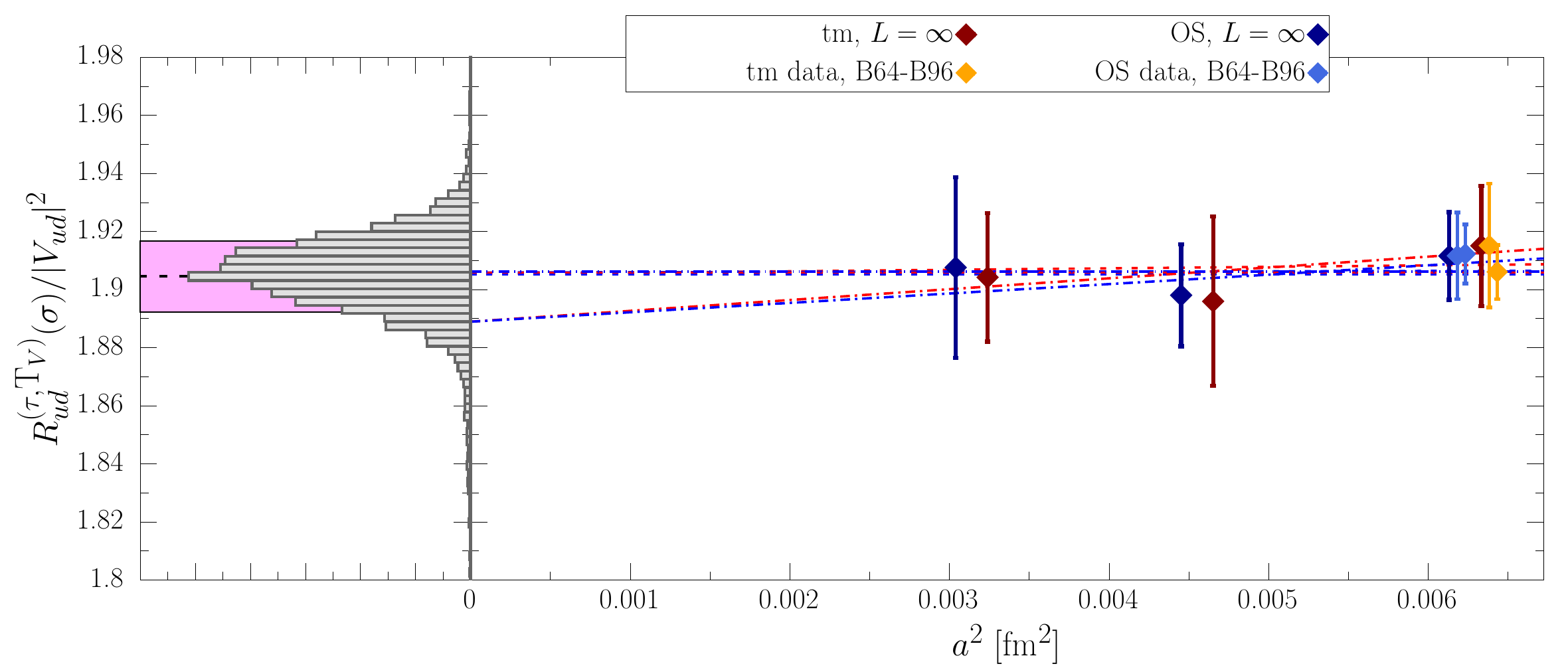}
    \includegraphics[scale=0.32]{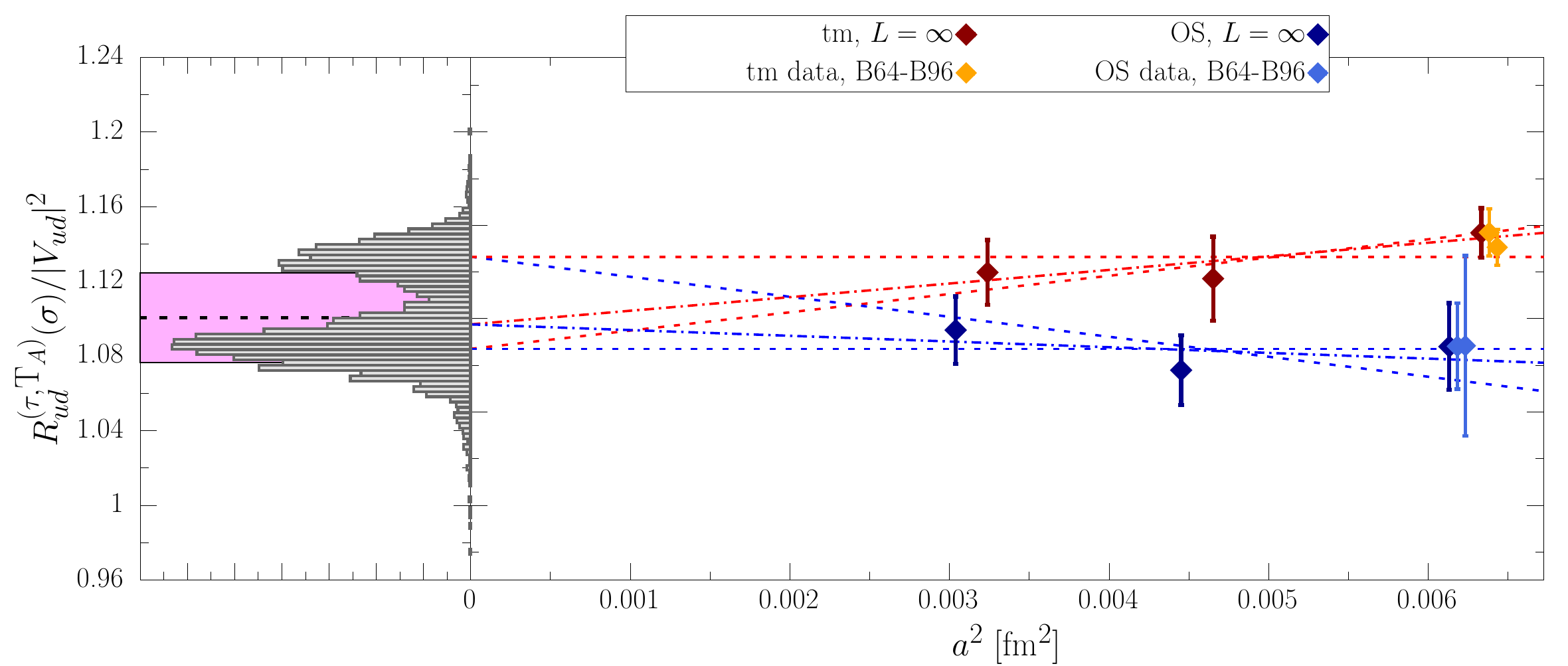}
    \includegraphics[scale=0.32]{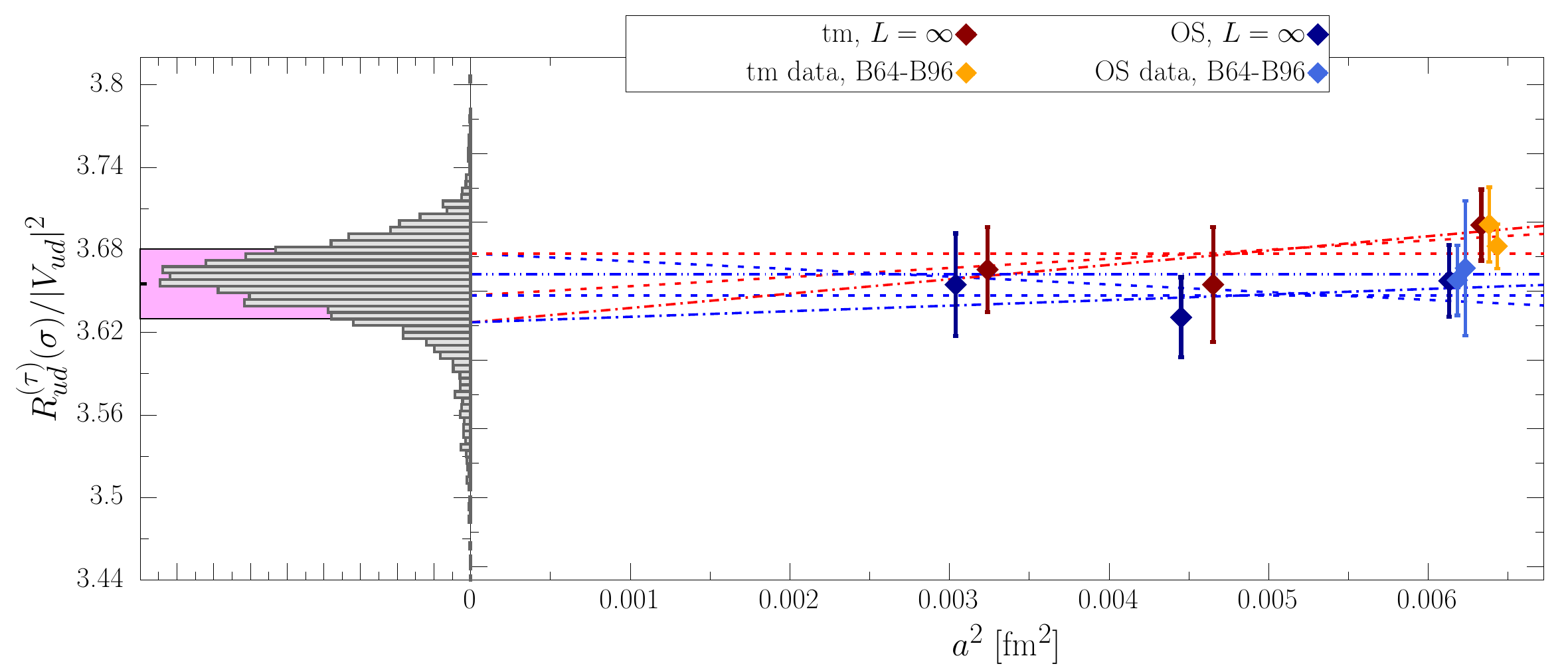}
    \caption{\small\it Continuum extrapolation of $R_{ud}^{(\tau,\mathrm{L})}(\sigma)$, $R_{ud}^{(\tau,\mathrm{T}_{V})}(\sigma)$, $R_{ud}^{(\tau,\mathrm{T}_{A})}(\sigma)$, and of the total $R_{ud}^{(\tau)}(\sigma)$ at a fixed value of $\sigma=0.004$ and for the case $\alpha=4$ and $r_{max}=4$. The red and blue data points correspond respectively to the lattice data obtained in the ``tm'' and ``OS'' regularization. The orange (``tm'') and light-blue (``OS'') data points at $a^{2}\simeq 0.006$ correspond to the raw-data obtained on the B64 and B96 ensembles, and are shown to highlight the size of the FSEs. The different red (for `tm'')  and blue (for `OS'') lines show a few of the fits obtained using a constant or linear Ansatz in $a^{2}$. The histograms shown in the left part of the panels of Fig.~\ref{fig:all_data} correspond to the distribution of the continuum extrapolated results obtained after applying the BMA procedure of Eq.~(\ref{eq:BMA}), using the weights in Eq.~(\ref{eq:AIC}). The magenta bands corresponds to our final determination.}
    \label{fig:all_data}
\end{figure}

In Fig.~\ref{fig:all_data} we illustrate the continuum fits we performed, by considering the case of the smallest value of $\sigma$ we employed, i.e. $\sigma=0.004$, and the reconstructed observables corresponding to the choice $\alpha=4$ and $r_{max}=4$ in the HLT method.
In the rightmost part of the panels of Fig.~\ref{fig:all_data}, the lattice data corresponding to the ``tm'' and ``OS'' regularization are shown in red and blue, respectively. The corresponding errorbars are already inclusive of the systematic errors due to FSEs and inexact kernel reconstruction. In the figure, to highlight the typical size of the FSEs, we also show the raw-data obtained on the B64 and B96 ensembles, which are indicated as orange and light-blue data-points, respectively for the ``tm'' and ``OS'' regularization. The different red (for `tm'')  and blue (for `OS'') lines show fits obtained using a constant or linear Ansatz in $a^{2}$. As it is clear from the figure, the cut-off effects are remarkably small, typically of the order of a few percent, and in some cases smaller than the combined statistical and systematic error of the data. Indeed, for many contributions we are able to obtain a very good $\chi^{2}/dof$ by fitting the data with a constant function.  Finally, in the leftmost part of the panels of Fig.~\ref{fig:all_data} we show the histograms corresponding to
the distribution of the continuum extrapolated results for the mean values of interest, 
as obtained by applying the BMA procedure of Eq.~(\ref{eq:BMA}) and using the weights in Eq.~(\ref{eq:AIC}). The magenta bands corresponds to our final determination. 

\begin{figure}
    \centering
    \hspace{-0.2cm}
    \includegraphics[scale=0.37]{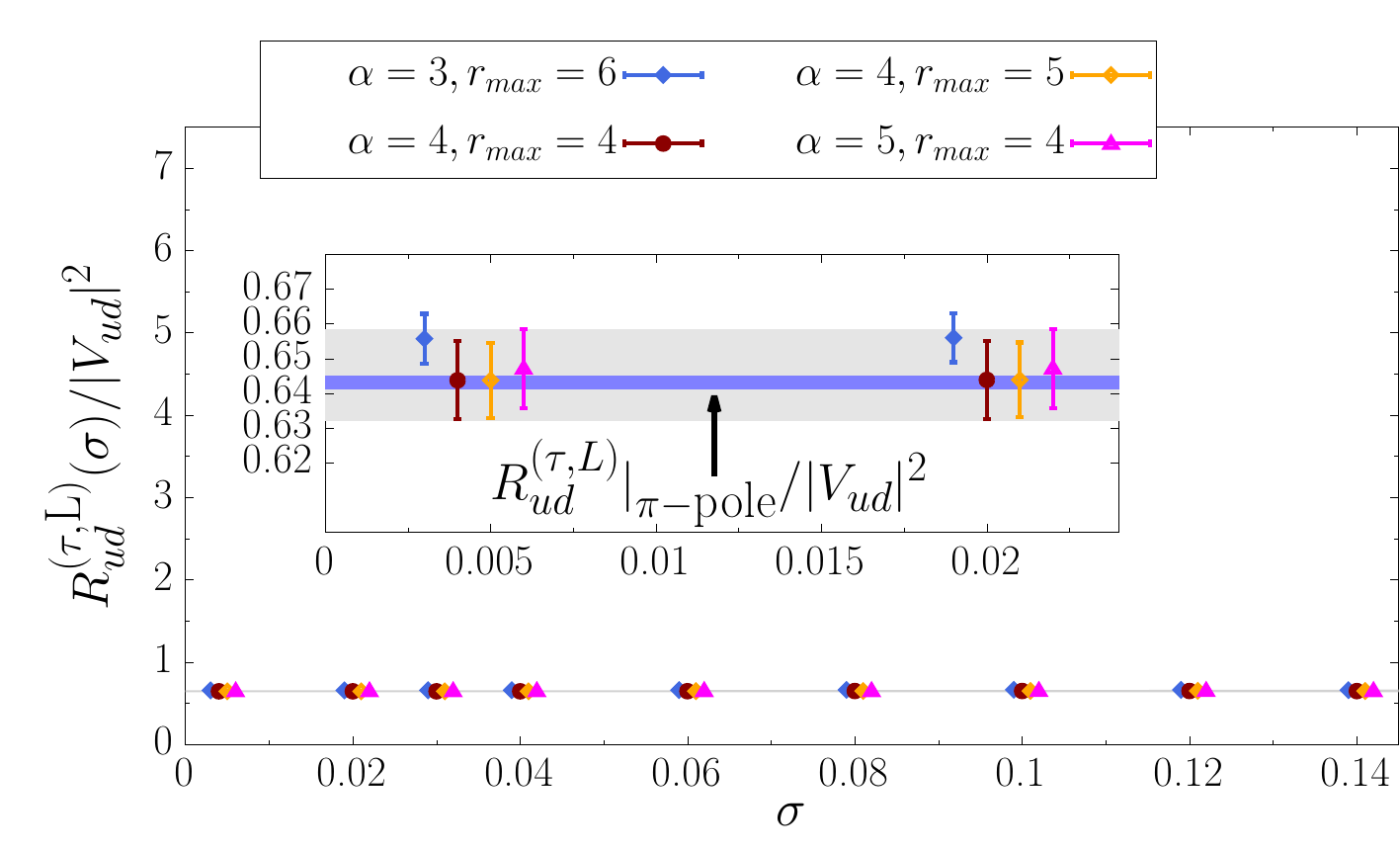}
    \includegraphics[scale=0.37]{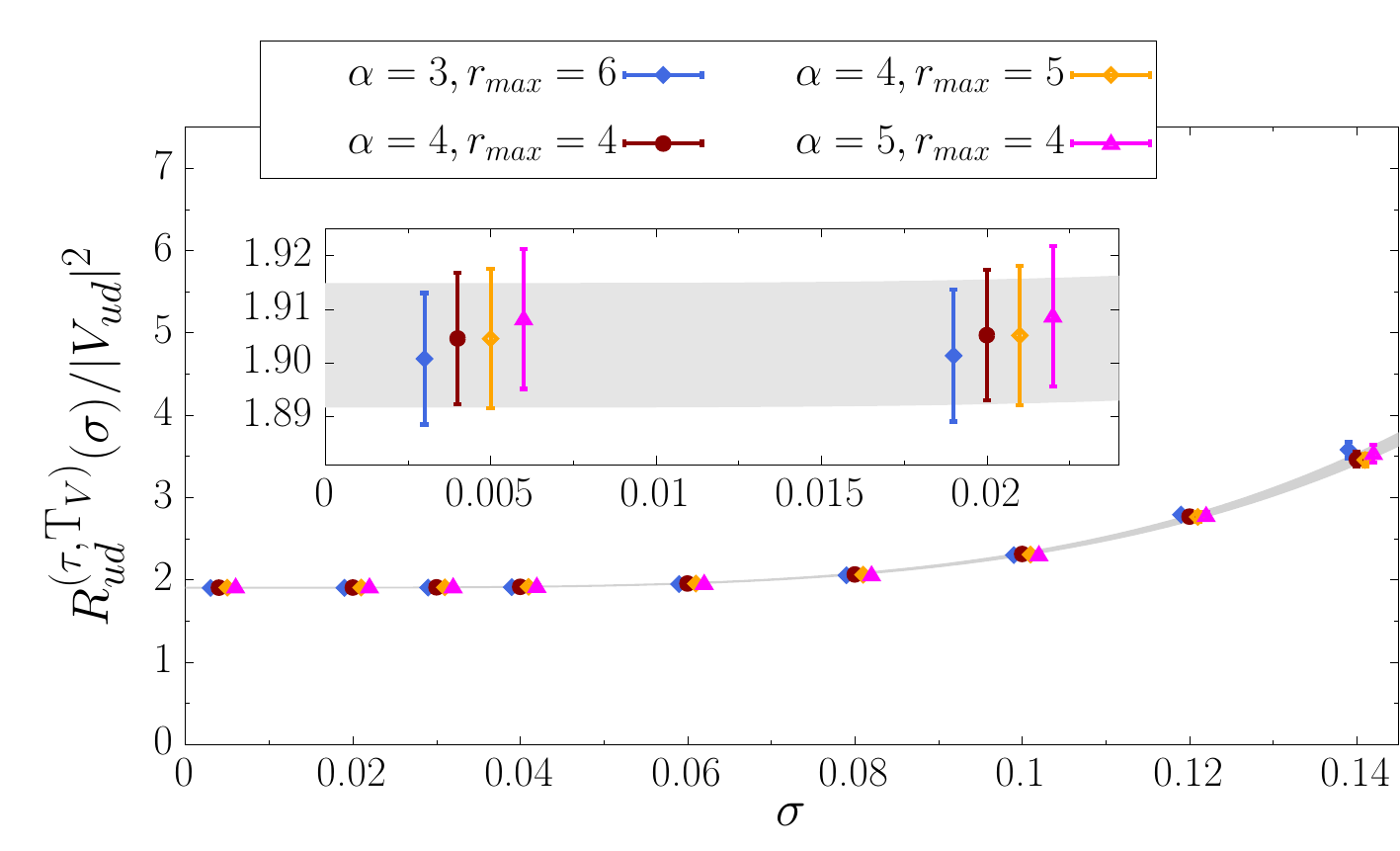}\\
    \includegraphics[scale=0.37]{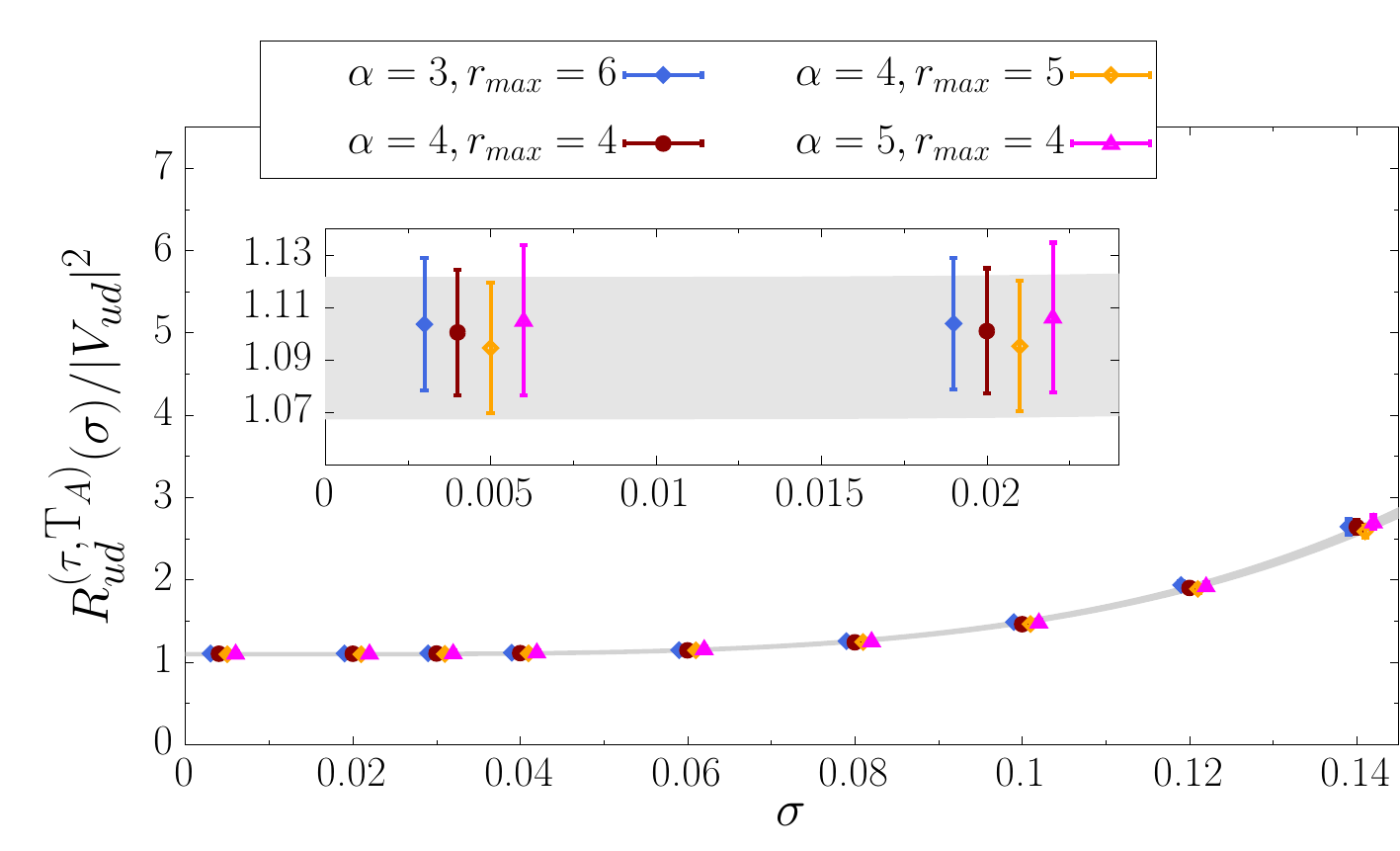}
    \includegraphics[scale=0.37]{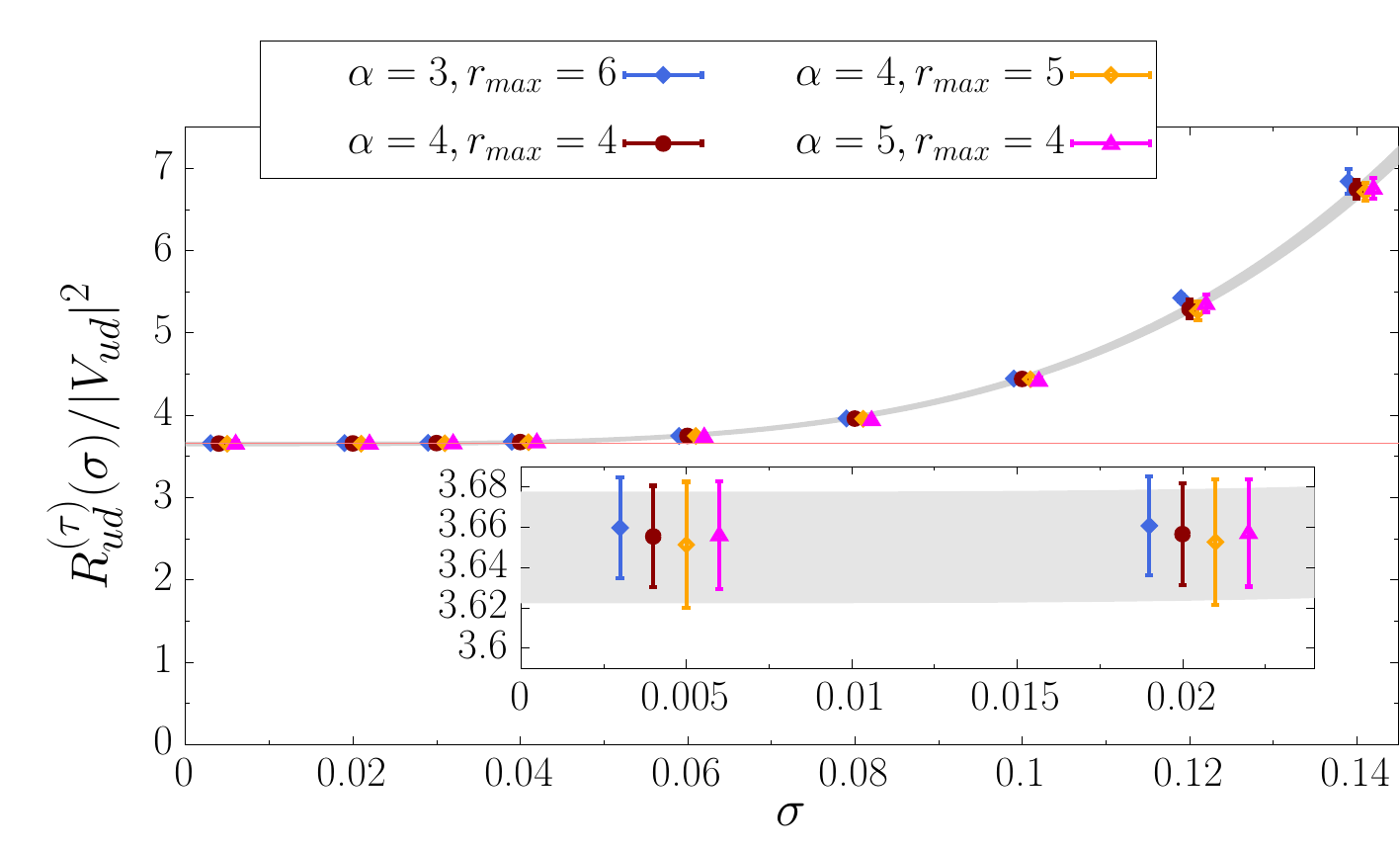}
    \caption{\small\it $\sigma-$dependence of the continuum extrapolated values of $R_{ud}^{(\tau,\mathrm{L})}(\sigma)$, $R_{ud}^{(\tau,\mathrm{T}_{A})}(\sigma)$, $R_{ud}^{(\tau,\mathrm{T}_{V})}(\sigma)$, and $R_{ud}^{(\tau)}(\sigma)$, as obtained after applying the BMA procedure of Eqs.~(\ref{eq:BMA})-(\ref{eq:AIC}). We show the results corresponding to different values of the algorithmic parameters $\alpha$ and $r_{max}$, which have been slightly shifted horizontally for better visibility. In the top-left panel, the blue band corresponds to the pion-pole prediction of Eq.~(\ref{eq:pion_pole}). Finally, the gray bands represent the result of the $\sigma\mapsto 0$ extrapolation using the Ansatz in Eq.~(\ref{eq:sigma4_ansatz}). }
    \label{fig:sigma_dep}
\end{figure}

Finally, let us now discuss the issue of the $\sigma\mapsto 0$ extrapolation. Under the assumption that the form factors $\rho_{\mathrm{T}}$ and $\rho_{\mathrm{L}}$ are regular at the end-point of the phase-space, i.e. for $E=m_{\tau}$, we show in Appendix \ref{app:B} that the corrections to the $\sigma\mapsto 0$ limit are even functions of $\sigma$, starting at $\order{\sigma^4}$, i.e.
\begin{align}
R_{ud}^{(\tau,\mathrm{I})}(\sigma)-R_{ud}^{(\tau,\mathrm{I})}= \mathcal{O}(\sigma^4)~.
\label{eq:sigmascaling}
\end{align}
The assumption of regularity is expected to hold true in the infinite-volume limit, where the form factors $\rho_{I}$ should be smooth functions of the energy around $E=m_{\tau}$. The result of Eq.~(\ref{eq:sigmascaling}) is important since it allows to carry out a controlled extrapolation to vanishing $\sigma$. In addition, since the leading order corrections are of order $\mathcal{O}(\sigma^{4})$, one expects a rather fast convergence of $R_{ud}^{(\tau)}(\sigma)$ towards $R_{ud}^{(\tau)}$. In Fig.~\ref{fig:sigma_dep}, we show as a function of the smearing parameter $\sigma$ our results for $R_{ud}^{(\tau,\mathrm{I})}(\sigma)$ after applying the BMA procedure of Eqs.~(\ref{eq:BMA})-(\ref{eq:AIC}). 
As it is clear from the figure, the results are remarkably stable under modification of the algorithmic parameters $\alpha$ and $r_{max}$. Moreover, for $\sigma < 0.04$, no dependence on $\sigma$ can be appreciated within errors, a remarkable finding that allows us to take the $\sigma\mapsto 0$ limit with confidence. For the longitudinal contribution $R_{ud}^{(\tau, \mathrm{L})}(\sigma)$ the dependence on $\sigma$ is practically absent over all range of $\sigma$ explored. This behaviour is somehow expected, since, as highlighted in the figure, $\rho_{\mathrm{L}}(E^{2})$ turns out to be dominated by the pion-pole contribution, which is given by
\begin{align}
\label{eq:pion_pole}
\rho_{\mathrm{L}}(E^{2})\big|_{\pi-{\rm pole}} = \pi \delta( E - m_{\pi})  \frac{f_{\pi}^{2}}{m_{\pi}} \quad \implies \quad R_{ud}^{(\tau,\mathrm{L})}\big|_{\pi-{\rm pole}} = 12\pi^{2} S_{EW} |V_{ud}|^{2} \frac{f_{\pi}^{2}}{m_{\tau}^{2}}\left( 1 - \frac{m_{\pi}^{2}}{m_{\tau}^{2}}\right)^{2} \simeq 0.643\,(2) |V_{ud} |^{2} ~,
\end{align}
and is thus less sensitive to the smearing, which mostly affects the behaviour of the kernel functions around $E \simeq m_{\tau}$, where the spectral function $\rho_{\mathrm{L}}(E^{2})$ is thus presumably small. 

In light of the asymptotic-expansion formula of Eq.~(\ref{eq:sigmascaling}), we have carried out the $\sigma\mapsto 0$ extrapolation employing the following linear Ansatz  in $\sigma^{4}$
\begin{align}
\label{eq:sigma4_ansatz}
R_{ud}^{(\tau,\mathrm{I})}(\sigma) = R_\mathrm{I} + A_\mathrm{I}\sigma^{4}~,
\end{align}
where $R_\mathrm{I}$ and $A_\mathrm{I}$ are free fit parameters. The extrapolation has been performed using our preferred analysis branch, i.e. the one with $\alpha=4$ and $r_{max}=4$, which leads to slightly smaller errors. We found that for all contributions the data up to $\sigma = 0.14$ are well described by the fit Ansatz, and stable within errors under removal of few data points at the largest $\sigma$-values
as well as under variation of the fit Ansatz by inclusion of a $\sigma^6$ term.
The result of the extrapolations, which basically coincide with the results obtained at the smallest value of $\sigma$ we simulated (i.e. $\sigma=0.004$),  are indicated by the gray bands in Fig.~(\ref{fig:sigma_dep}).

Our main results are summarized in Tab.~\ref{tab:final_res}, where we provide also the axial  contribution $R_{ud}^{(\tau,A)}$ given by
\begin{align}
R_{ud}^{(\tau,A)} \equiv R_{ud}^{(\tau, \mathrm{T}_{A})}+ R_{ud}^{(\tau, \mathrm{L})}~,
\end{align}
while, as already mentioned, the vector contribution $R_{ud}^{(\tau, V)}$ coincides, in the limit of degenerate up and down quarks in which we work, with $R_{ud}^{(\tau, \mathrm{T}_{V})}$. The overall accuracy that we obtained for the different contributions is very good, typically of order $\mathcal{O}(1\%)$ or smaller, and motivates us to undertake the task of computing the leading isospin breaking corrections for this quantity,
which, given that $\alpha_{\rm em} \simeq (m_{d}-m_{u})/\Lambda_{QCD}\simeq \mathcal{O}(1\%)$, are expected to be of the same order of magnitude
as our present total (statistical and systematic) uncertainty.

\begin{figure}
    \centering
    \includegraphics[scale=0.39]{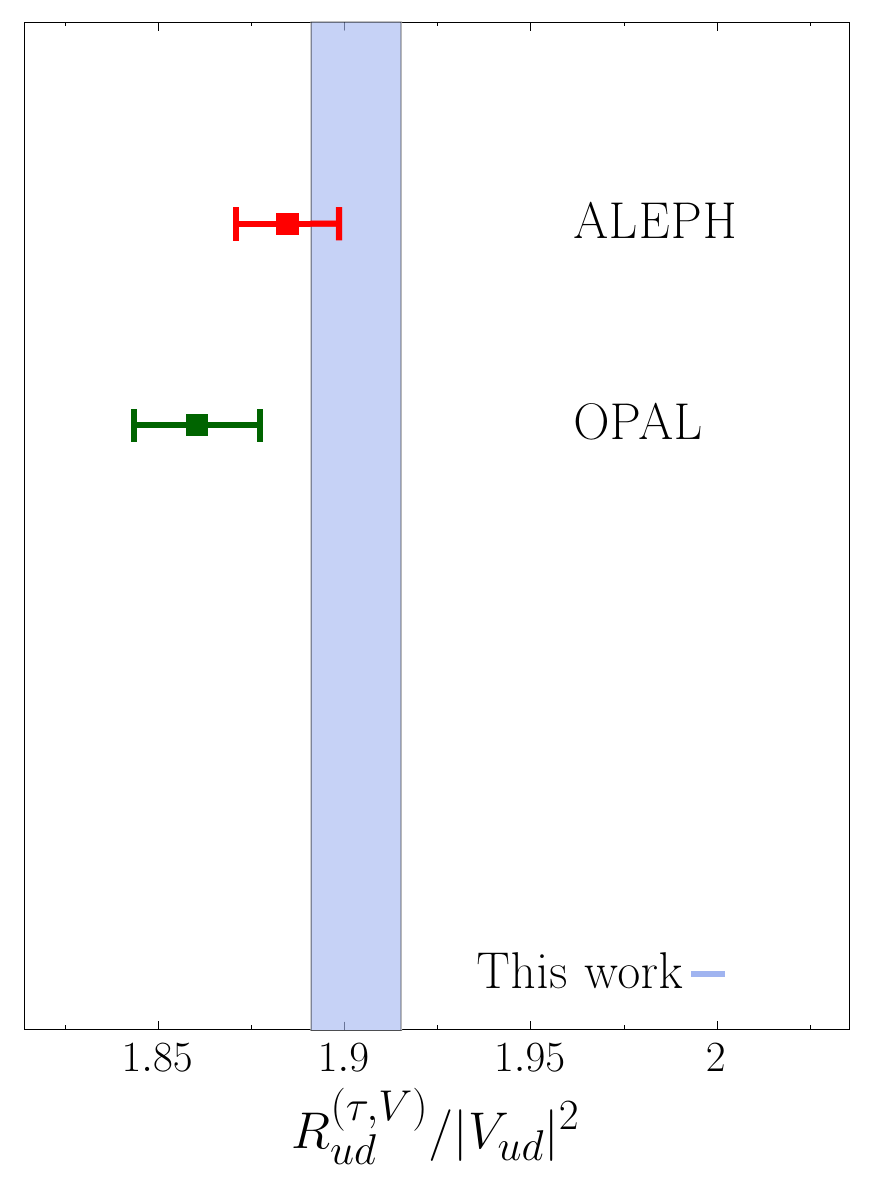}
    \includegraphics[scale=0.39]{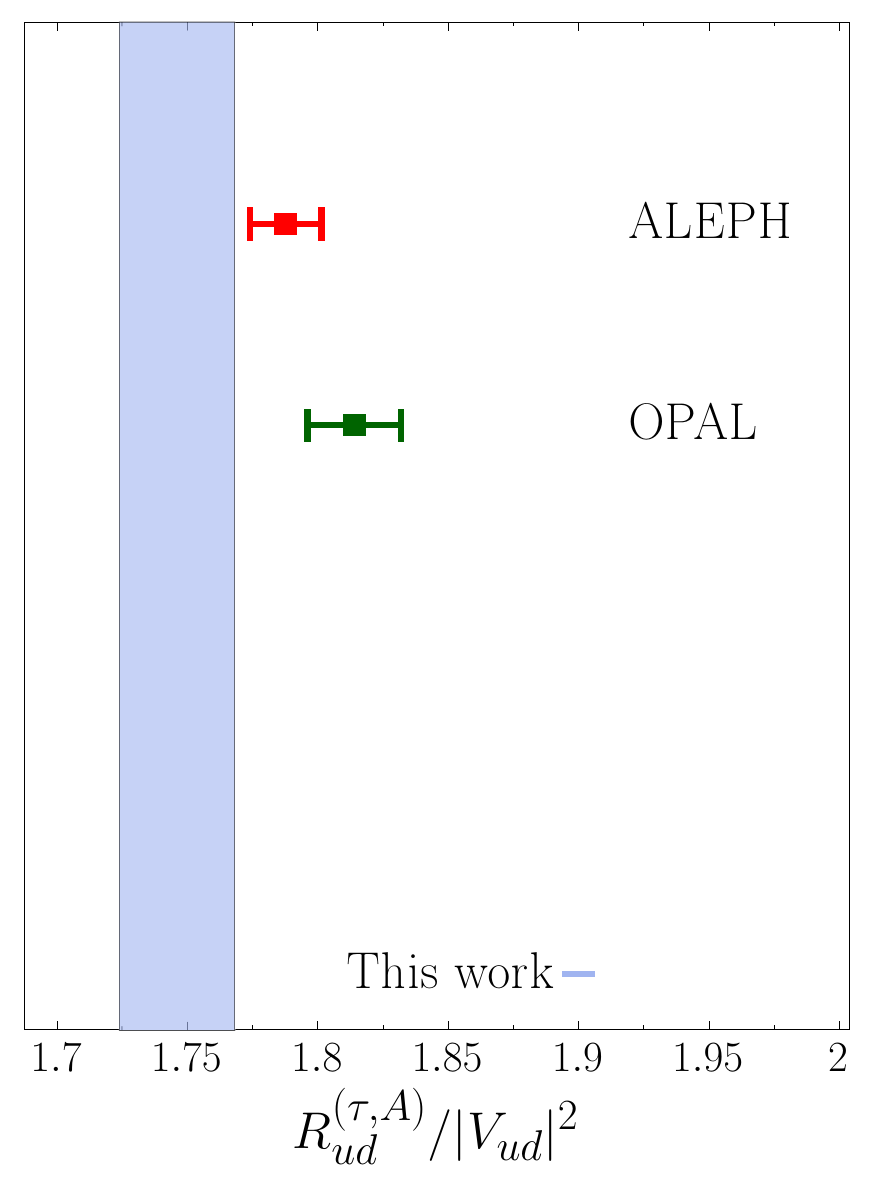}
    \includegraphics[scale=0.39]{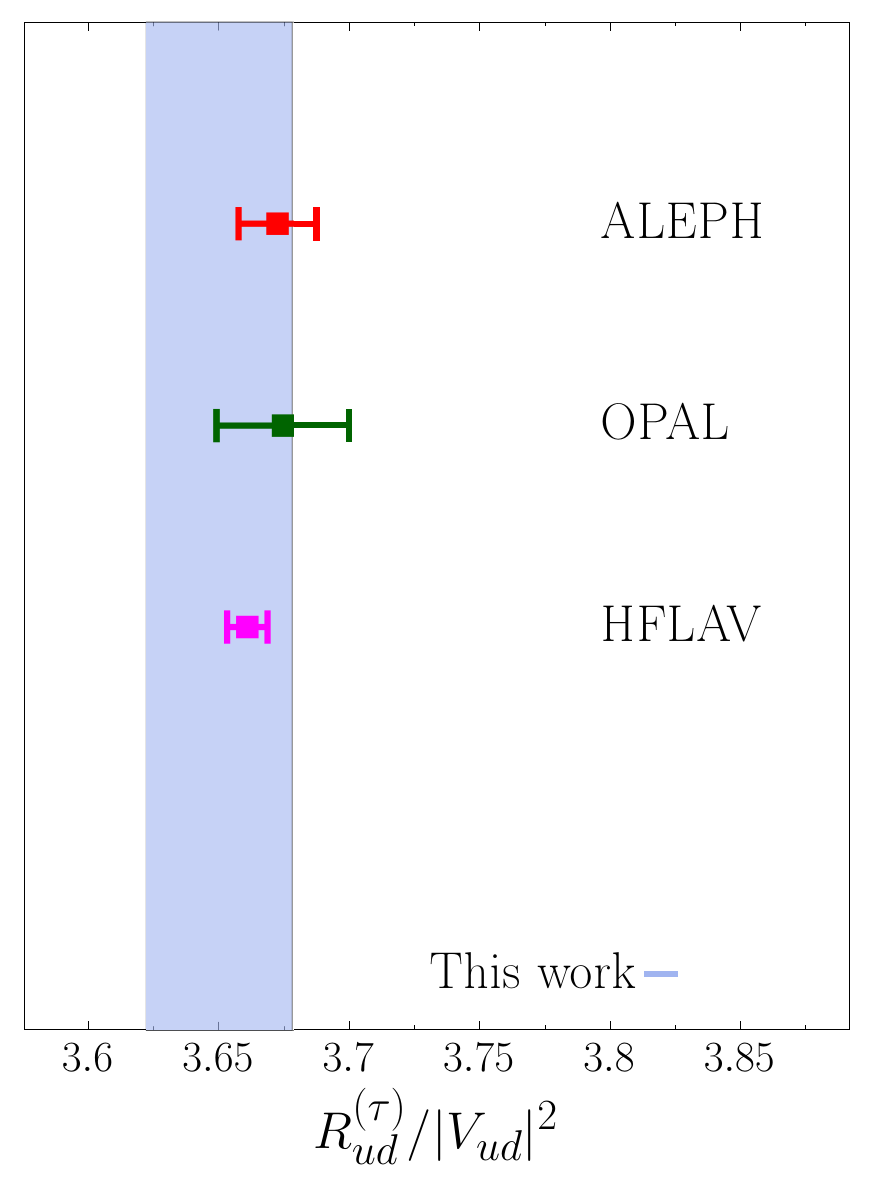}
    \caption{\small\it Comparison between the lattice results obtained in this work and the experimental measurements by the ALEPH~\cite{ALEPH:2005qgp} and OPAL~\cite{OPAL:1998rrm} Collaboration. We show, from the left to the right, the vector contribution $R_{ud}^{(\tau, V)}$, the axial contribution $R_{ud}^{(\tau,A)}$, and the total $R_{ud}^{(\tau)}$. In the rightmost panel, the data point in magenta corresponds to the HFLAV average~\cite{HeavyFlavorAveragingGroup:2022wzx} of the experimental results. For this comparison, we divided the experimental results for $R_{ud}^{(\tau,V)}, R_{ud}^{(\tau,A)}$ and $R_{ud}^{(\tau)}$ by $|V_{ud}|^{2}$ using the value~\cite{FlavourLatticeAveragingGroupFLAG:2021npn} $|V_{ud}| = 0.97373~(31)$.}
    \label{fig:comparison_ALEPH}
\end{figure}
\begin{table}[]
    \centering
    \begin{tabular}{|c|| c | c | c | c | c | c  ||  }
    \hline
       & ~ $\mathrm{L}$ ~ & ~ $\mathrm{T}_{A}$ ~ & $\mathrm{T}_{V}$ ~ & ~ $\mathrm{T}$ ~ & ~ $A$ ~ & ~ tot ~ \\ 
       \hline 
       This work &  ~ $0.645~(13)$ ~ & ~ $1.094~(27)$ ~ & ~ $1.903~(12)$ ~ & ~ $2.996~(33)$ ~ & ~ $1.746~(22)$ ~ & ~ $3.650~(28)$ ~   \\ 
       \hline
       \hline 
    \end{tabular}
    \caption{\it\small The main results obtained in this work. We give, in order from left to right, our predictions for $R_{ud}^{(\tau, W)}/|V_{ud}|^{2}$ with $W= \mathrm{L}, \mathrm{T}_{A}, \mathrm{T}_{V}, \mathrm{T}, A$. The value in the last column corresponds to the total $R_{ud}^{(\tau)}/|V_{ud}|^{2}$.   }
    \label{tab:final_res}
\end{table}

In Fig.~\ref{fig:comparison_ALEPH} we compare our lattice results, with the experimental data from the ALEPH~\cite{ALEPH:2005qgp} and OPAL~\cite{OPAL:1998rrm} collaborations. As the figure shows, we find for $R_{ud}^{(\tau)}/|V_{ud}|^{2}$ a good agreement between our results and the experimental measurements. For the vector and axial contributions, we observe some differences with respect to the experimental data, which are more pronounced in the axial channel (at the level of $1.6\sigma$ and $2.4\sigma$ if we compare with the ALEPH and OPAL results, respectively). This difference is however at the level of a few percent and might well be attributed to the missing isospin breaking contributions and/or to statistical fluctuations. In the rightmost panel of Fig.~\ref{fig:comparison_ALEPH} we  report the HFLAV average~\cite{HeavyFlavorAveragingGroup:2022wzx}
\begin{align}
\label{eq:HFLAV_average}
\frac{1}{|V_{ud}|^{2}} R_{ud}^{(\tau)}({\rm HFLAV})= 3.660~(8)~,
\end{align}
obtained by combining, for each exclusive decay mode, the results of several experiments (see Ref.~\cite{HeavyFlavorAveragingGroup:2022wzx} and references therein for details). This value compares very well with our theoretical prediction
\begin{align}
\frac{1}{|V_{ud}|^{2}}R_{ud}^{(\tau)} = 3.650~(28)~.
\end{align}
For the comparison shown in Fig.~\ref{fig:comparison_ALEPH} we used the most precise determination of $|V_{ud}|$ which comes from the study of the superallowed $0^+ \to 0^+$ nuclear
beta decays, and it is given by $|V_{ud}| = 0.97373~(31)~$~\cite{FlavourLatticeAveragingGroupFLAG:2021npn}.

In alternative to the comparison shown in Fig.~\ref{fig:comparison_ALEPH},  our results for $R_{ud}^{(\tau)}/|V_{ud}|^{2}$ can be used to determine $|V_{ud}|$ using the HFLAV average $R_{ud}^{(\tau)}({\rm HFLAV}) = 3.471~(7)$.
The analysis yields
\begin{align}
|V_{ud}| = 0.9752~(39)~,
\end{align}
in agreement with the determination from the superallowed nuclear beta decays. Exploiting the unitarity of the CKM matrix, we also obtain for $|V_{us}|$ the estimate
\begin{align}
|V_{us}| = \sqrt{ 1 - |V_{ud}|^{2} - |V_{ub}|^{2}} \simeq \sqrt{ 1 - |V_{ud}|^{2}} = 0.221~(17)~.
\end{align}
Finally, another interesting quantity to provide is the difference between the vector and axial contribution, normalized over the total ratio $R_{ud}^{(\tau)}$, namely
\begin{align}
\label{eq:delta_ratio}
\Delta^{(\tau)}_{V-A} \equiv \frac{R_{ud}^{(\tau,V)}-R_{ud}^{(\tau,A)}}{R_{ud}^{(\tau)}}~.
\end{align}
This quantity, which is independent from $|V_{ud}|$,   vanishes to any given order of the perturbative expansion (if we neglect
light-quark mass effects), and is thus very sensitive to non-perturbative physics contributions. The ALEPH and OPAL collaborations quote the value
\begin{align}
\label{eq:ratio_exp}
\Delta^{(\tau)}_{V-A}({\rm ALEPH}) = 0.026~(7)~,\qquad \Delta^{(\tau)}_{V-A}({\rm OPAL}) = 0.013~(7)~,
\end{align}
which can be compared with our determination
\begin{align}
\label{eq:ratio_theo}
\Delta_{V-A}^{(\tau)} = 0.042~(5)~.
\end{align}
Due to the large cancellation in $\Delta_{V-A}^{(\tau)}$ between the vector and axial contribution, the resulting uncertainties for this quantity are, at present, quite sizable: they are of order $\mathcal{O}(10\%)$ in the case of our theoretical prediction, and of order $\mathcal{O}(25-30\%)$ and $\mathcal{O}(50\%)$  for the ALEPH and OPAL results, respectively.
The difference between our prediction and the experimental values in Eq.~(\ref{eq:ratio_exp}) is at the level of $1.9\sigma$ and $3.5\sigma$, if we compare with the ALEPH and OPAL results, respectively. However, for this quantity, which is obtained after a strong cancellation between the vector and axial contribution, the relative impact of the missing isospin breaking corrections might be stronger, and for this reason, at present, we cannot claim any discrepancy between the theoretical prediction and the experimental value of $\Delta_{V-A}^{(\tau)}$. In the future, it will be interesting to see whether this difference increases or vanishes once isospin breaking corrections are included, and both experimental and theoretical uncertainties reduced.

\section{Conclusions}
\label{sec:conclusions}
In this paper we have presented, for the first time, a first-principle and fully non-perturbative lattice QCD determination of the inclusive decay rate of the $\tau$-lepton. The method relies on the HLT method of Ref.~\cite{Hansen:2019idp}, which allows to evaluate the energy-integral of spectral density weighted by smooth analytic kernel functions. In this first numerical study, we have evaluated the (semi-)inclusive decay rate $\tau\to X_{ud}\nu_{\tau}$, where $X_{ud}$ is a generic hadronic state with $\bar{u}d$ flavor quantum numbers, in isospin-symmetric QCD, i.e. by neglecting strong and electromagnetic isospin-breaking corrections. The method we propose does not rely on OPE or perturbative approximations and proceeds by regularizing the expression for the ratio $R_{ud}^{(\tau)}$ of decay rates in Eqs.~(\ref{Rtaudef}) and~(\ref{Rtau}), which involves an integral over the spectral form factors $\rho_{T}(s)$ and $\rho_{L}(s)$, by introducing a non-zero smearing parameter $\sigma$. The resulting regularized ratio $R_{ud}^{(\tau)}(\sigma)$ (see Eq.~(\ref{eq:R_sigma})) can be then targeted by the HLT method, and the limit of vanishing $\sigma$, which must be considered in order to recover the physical ratio, is then taken after the appropriate infinite-volume and continuum-limit extrapolations. 

We have shown that the $\sigma\mapsto 0$ extrapolation can be handled in a completely controlled way, thanks to the analytical result  derived in Appendix~\ref{app:B} for the corrections to the vanishing $\sigma$ limit of $R_{ud}^{(\tau)}(\sigma)$. For this quantity, the corrections start at $\mathcal{O}(\sigma^{4})$, and numerically we have found clear evidences of the presence of such behaviour. Furthermore, for $\sigma < 0.04$, we are already in the region where the $\mathcal{O}(\sigma^{4})$ corrections are negligibly small, and no dependence on $\sigma$ can be observed within our current uncertainties. These findings allow us to take the $\sigma\mapsto 0$ limit in full confidence. 

Making use of the last generation of ensembles produced in isospin symmetric QCD by the ETMC at three lattice spacings, two volumes, and with $N_{f}=2+1+1$ quark flavours at physical mass values (see Tab.~\ref{tab:simudetails} and Refs.~\cite{ExtendedTwistedMass:2021gbo,ExtendedTwistedMass:2021qui,ExtendedTwistedMass:2022jpw,Alexandrou:2018egz}), we have determined both the total ratio $R_{ud}^{(\tau)}$ and its individual contributions coming from the longitudinal or transverse part of the spectral density tensor in Eq.~(\ref{rhoLT}), and from the axial or vector part of the weak current. We obtained a satisfactory accuracy for all contributions; our final errors are typically of order of $\mathcal{O}(1\%)$, or smaller. These findings motivate us to compute the only missing piece in our calculation, namely the isospin breaking corrections, which are parametrically of order $\alpha_{em} \simeq (m_{d}-m_{u})/\Lambda_{QCD}\simeq \mathcal{O}(1\%)$, and thus of the same order of magnitude as our current (statistical and systematic) uncertainties. 

From the phenomenological viewpoint, our theoretical prediction for $R_{ud}^{(\tau)}/|V_{ud}|^{2}$ can be compared with the experimental results for $R_{ud}^{(\tau)}$, by using an independent determination of $|V_{ud}|$ (e.g. the precise value obtained from the superallowed $\beta$ decays~\cite{FlavourLatticeAveragingGroupFLAG:2021npn}).  For the total $R_{ud}^{(\tau)}$, our determination compares very well with the HFLAV average of the experimental results~\cite{HeavyFlavorAveragingGroup:2022wzx}. The separate vector and axial contributions, $R_{ud}^{(\tau, V)}/|V_{ud}|^{2}$ and $R_{ud}^{(\tau,A)}/|V_{ud}|^{2}$, 
can be instead compared with the experimental determination provided by the ALEPH~\cite{ALEPH:2005qgp} and OPAL~\cite{OPAL:1998rrm} collaborations. In the axial channel, we observe some difference with  the experimental results, which is at the level of $1.6\sigma$ and $2.4\sigma$, if we compare with the ALEPH and OPAL measurements, respectively. Clearly, such difference might be due to the missing isospin breaking corrections, as well as to statistical fluctuations. We also provided our prediction for the $|V_{ud}|$-independent ratio $\Delta_{V-A}^{(\tau)}$ in Eq.~(\ref{eq:delta_ratio}), which vanishes to any given order of the perturbative expansion in massless QCD, and is thus very sensitive to the non-perturbative contributions. Also for this quantity we observe some difference with respect to the experimental data, in particular, again, with the OPAL results (at the level of $3.5\sigma$). Once isospin breaking corrections will be included, it will be extremely interesting to check whether this difference increases or disappears. 

Finally, we would like to briefly discuss the prospects for the extraction of the CKM matrix elements from inclusive $\tau$ decays. In the $ud$-flavour channel, as already mentioned, the current most precise determination of $|V_{ud}|$ comes from the superallowed $\beta$ decays, and has a striking uncertainty of $\mathcal{O}(0.03\%)$. In order to match such level of accuracy, the experimental uncertainty on $R_{ud}^{(\tau)}$ should be reduced by a factor $\simeq 3$, while from the theory side, the lattice QCD (+QED) errors on $R_{ud}^{(\tau)}/|V_{ud}|^{2}$ should be reduced by approximately one order of magnitude. While matching such level of accuracy seems unfeasible in the short term, having an independent and gradually more precise determination of $|V_{ud}|$ is important, since it allows to monitor whether the (currently existing) agreement between inclusive and exclusive determinations of $|V_{ud}|$ remains, as the errors get reduced.

The approach we discussed here considering the $ud$-flavour channel can be immediately applied to the $us$-flavour channel as well. In this case, having a precise determination of $|V_{us}|$ from the (semi-)inclusive decay $\tau \to X_{us}\nu_{\tau}$, would be even more interesting from a phenomenological point of view. Indeed, the typical accuracy of the current most precise determinations of $|V_{us}|$ is approximately in the range $0.3-0.7\%$, which could be matched by the precision of the lattice calculation of $R_{us}^{(\tau)}/|V_{us}|^{2}$. From the experimental side instead, the present uncertainty on $R_{us}^{(\tau)}$ is, according to the HFLAV review~\cite{HeavyFlavorAveragingGroup:2022wzx}, of about $1.6\%$, which would allow for an extraction of $|V_{us}|$ at the $0.8\%$ precision level. This is not yet at the level of precision already reached by other methods, but it could be expected that the experimental accuracy will improve over the next years. In this respect, we believe that providing  a first-principle and fully non-perturbative determination of $R_{us}^{(\tau)}/|V_{us}|^{2}$ will encourage the experimentalists to work in this direction.

\section{Acknowledgements}
We thank all the members of the ETMC for the most enjoyable collaboration. We acknowledge CINECA for the provision of CPU time under the specific
initiative INFN-LQCD123 and IscrB\_S-EPIC. F.S. G.G and S.S. are supported by the Italian Ministry
of University and Research (MIUR) under grant PRIN20172LNEEZ. F.S. and G.G are supported by
INFN under GRANT73/CALAT. F.S. is supported by ICSC – Centro Nazionale di Ricerca in High Performance
Computing, Big Data and Quantum Computing, funded by European Union –
NextGenerationEU.

\newpage
\appendix

\section{Inclusive decay rate, optical theorem and the Cutkosky rule}
\label{app:A}

\begin{figure}[t]
    \centering
    \includegraphics[scale=0.3]{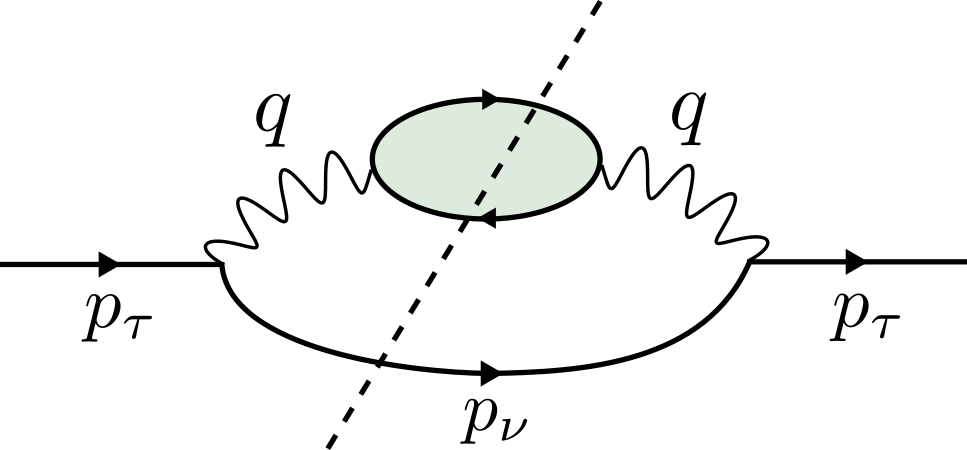}
    \caption{\small\it Feynman diagram representing, through the optical theorem, the inclusive $\tau$ decay rate.}
    \label{fig:optical}
\end{figure}

In this appendix we present an alternative derivation of the expression for the inclusive decay rate $\Gamma_{ud}^{(\tau)}$ of Eq.\,\eqref{Gamma} based on the optical theorem and the Cutkosky rule (the relevant Feynman diagram and the corresponding cut are shown in Fig.\,\ref{fig:optical}). 

The optical theorem relates the inclusive decay rate $\Gamma^{(\tau)}$ to the forward matrix element of the transition matrix $T$ (with $S = 1 + i\,T$),
\be
\label{optical}
    \Gamma^{(\tau)} = \frac{1}{2\, m_\tau} \, 2 \Im\,T_{\tau\tau} \ ,
\ee
where $T_{\tau\tau} = \langle \tau\, | T | \, \tau \rangle$.  
%%%%%%%%%%%%%%%%%%%%%%%%%%%%%%%%%%%%%%%%%%%
%%%%%%%%%%%%%%%%%%%%%%%%%%%%%%%%%%%%%%%%%%%
At the leading order in the Fermi effective theory, and for intermediate hadronic states with flavor quantum number $\bar u d$, the transition amplitude is expressed by
\be
\label{iT}
    i\, T_{\tau\tau} = \left(\frac{G_F}{\sqrt{2}} \right)^2 |V_{ud}|^2 \int \frac{d^4 p_\nu}{(2\pi)^4} \, \frac{i}{p_\nu^2 + i \varepsilon} \, L_{\alpha\beta}(p_{\tau}, p_\nu) \ i\, \Pi_{ud}^{\alpha\beta}(q) \ , \qquad q = p_{\tau}-p_\nu \ ,
\ee
where $L_{\alpha\beta}(p_{\tau}, p_\nu)$ is the leptonic tensor of Eq.\,\eqref{leptonic} and $\Pi_{ud}^{\alpha\beta}(q)$ is the hadronic vacuum polarization tensor,
\be
\label{HVP}
    \Pi_{ud}^{\alpha\beta}(q) = i \, \int d^4x \, e^{iqx} \, \langle 0 | T \left( J_{ud}^\alpha(x) \, J_{ud}^\beta(0)^\dagger \right) |0\rangle \ .
\ee

In order to evaluate the imaginary part of the transition amplitude, it is useful to consider for $\Pi_{ud}^{\alpha\beta}(q)$ its K\"{a}llen-Lehmann representation, which reads
\be
\label{KL}
-i \, \Pi_{ud}^{\alpha\beta}(q) = \int_0^\infty \frac{dM^2} {2\pi}\, \frac{i}{q^2 - M^2 + i \varepsilon} \ \rho_{ud}^{\alpha\beta}(q_M) \ ,
\ee
where $\rho_{ud}^{\alpha\beta}(q)$ is the spectral density of Eq.\,\eqref{rho} and $q_M =(\sqrt{M^2+\bs{q}^2},\, \bs{q})$ (so that $q_M^2 = M^2$). By inserting Eq.\,\eqref{KL} into Eq.\,\eqref{iT}, one obtains
\be
\label{iT2}
    i\, T_{\tau\tau} = 
    \left(\frac{G_F}{\sqrt{2}} \right)^2 |V_{ud}|^2 \int \frac{d^4 p_\nu}{(2\pi)^4} \, \frac{L_{\alpha\beta}(p_{\tau}, p_\nu) }{p_\nu^2 + i \varepsilon} \,
    \int_0^\infty \frac{dM^2} {2\pi}\, \frac{\rho_{ud}^{\alpha\beta}(q_M) }{q^2 - M^2 + i \varepsilon} \ .
\ee

The imaginary part of the transition amplitude can now be evaluated in terms of the discontinuity of the amplitude across the real axis, in the complex-$p^2$ plane, which is related to $\Im(T)$ by
\be
\textrm{Disc}[T(p^2)] = T(p^2+i \varepsilon) - T(p^2-i \varepsilon) = 2\, i \Im[T(p^2)] \ .
\ee
In turn, the discontinuity of $T$ can be evaluated using the Cutkosky rule, which requires replacing each propagator in the cuts of a diagram with its discontinuity across the real axis, namely
\be
\label{Cut}
     \frac{i}{p^2 -m^2 +i \varepsilon} \ \longrightarrow \ \textrm{Disc}\left[ \frac{i}{p^2 -m^2 +i \varepsilon} \right] = 2\pi \, \delta^+ (p^2-m^2) \ ,
\ee
where $\delta^+ (p^2-m^2) = \delta (p^2-m^2)\, \theta(p^0)$. Applied to the the K\"{a}llen-Lehmann representation of $\Pi_{ud}^{\alpha\beta}(q)$ in Eq.\,\eqref{KL}, the Cutkosky rule leads to 
\be
\label{KLcut}
2 \Im[\Pi_{ud}^{\alpha\beta}(q)] = \textrm{Disc}[-i \, \Pi_{ud}^{\alpha\beta}(q)] = \int_0^\infty dM^2 \, \delta^+ (q^2-M^2) \, \rho_{ud}^{\alpha\beta}(q_M) = \rho_{ud}^{\alpha\beta}(q) \ ,
\ee
which shows that the spectral density is just twice the imaginary part of the polarization tensor $\Pi_{ud}^{\alpha\beta}(q)$. Applying instead the Cutkosky rule to the neutrino's propagator in  Eq.\,\eqref{iT2}, the integral over the neutrino 4-momentum is reduced to an integral over the phase-space of the external particle,
\be
\label{nucut}
    \int \frac{d^4 p_\nu}{(2\pi)^4} \frac{i}{p_\nu^2 + i \varepsilon} \ \longrightarrow \ 
    \int \frac{d^4 p_\nu}{(2\pi)^4} \ 2\pi \, \delta^+(p_\nu^2) = \int \frac{d^3 p_\nu}{(2\pi)^3 \, 2 E_\nu} \ .
\ee
Thus, from the optical theorem of Eq.\,\eqref{optical}, and using Eqs.\,\eqref{iT2}, \eqref{KLcut} and \eqref{nucut}, one arrives at
\be
\label{GammaApp}
    \Gamma_{ud}^{(\tau)} =\frac{G_F^2 \, |V_{ud}|^2}{4\, m_\tau} \int \frac{d^3 p_\nu}{(2\pi)^3 \, 2 E_\nu} \ L_{\alpha\beta}(p_{\tau}, p_\nu) \ \rho_{ud}^{\alpha\beta}(q) \ ,   \qquad q = p_{\tau} - p_\nu \ ,
\ee
in agreement with the result obtained in Eq.\,\eqref{Gamma}.

\section{Corrections to the $\sigma \mapsto 0$ spectral densities reconstruction}
\label{app:B}

In this appendix we derive the corrections to the $\sigma \mapsto 0$ limit of the inclusive ratio $R_{ud}^{(\tau)}(\sigma)$ of Eq.~(\ref{eq:R_sigma}). To this end we start by rewriting Eq.~(\ref{eq:R_sigma}) as follows  
\begin{align}
R_{ud}^{(\tau)}(\sigma) = \int_{0}^{+\infty}\dd x\, (1-x)^2\, \Tilde{\rho}(x)\,\Theta_\sigma\left(1-x\right)\,,
\end{align}
where $x=E/m_\tau$ and where we have defined
\begin{align}
\Tilde{\rho}\left(x\right)
=
12 \pi \, S_{EW} \, |V_{ud}|^2\,  x(1+x)^2
\left[ (1+2x^2)\rho_{\mathrm{T}}(m_\tau^2 x^2) + \rho_{\mathrm{L}}(m_\tau^2 x^2)\right]\;.
\end{align}
By noticing the properties
\begin{align}
\Theta_\sigma(x) = \Theta_1\left(\frac{x}{\sigma}\right)\;,
\qquad\quad
\Theta_1(x) + \Theta_1(-x) = 1\,, 
\qquad\quad
x^p\,\partial^q_x\,\left[1-\Theta_1(x) \right]
\stackrel{x\mapsto \infty}{=}  \order{e^{-x}}\quad \forall \;p,q\in \mathbb{N}\,,
\label{eq:thetaprops}
\end{align}
satisfied by the smeared theta-function $\Theta_\sigma(x)$, see Eq.~(\ref{eq:sigmoid}), 
the corrections to the $\sigma \mapsto 0$ result 
\begin{align}
\Delta R_{ud}^{(\tau)}(\sigma)
= R_{ud}^{(\tau)}(\sigma) - R_{ud}^{(\tau)}
\end{align}
can be written as
\begin{align}
\Delta R_{ud}^{(\tau)}(\sigma) 
&=
\int_{0}^{\infty}\dd x\,\left\{\Theta_\sigma\left(1-x\right) - \theta(1-x)\right\}(1-x)^2\Tilde{\rho}(x)
\nonumber \\[8pt]
&=
\int_{0}^{\infty}\dd x\,\left\{\Theta_1\left(\frac{1-x}{\sigma}\right) - \theta\left(\frac{1-x}{\sigma}\right)\right\}(1-x)^2\Tilde{\rho}(x)
\nonumber \\[8pt]
&=
-\sigma^3\int_{-\infty}^{\frac{1}{\sigma}}\dd y\,\left\{
\theta\left(y\right) -\Theta_1\left(y\right)\right\}\,
y^2\Tilde{\rho}(1-\sigma y)\;.
\label{eq:DeltaRstep1}
\end{align}
In the previous expressions we have used the fact that $\theta(x)=\theta(x/\sigma)$ for $\sigma>0$ and made the change of variables $y=(1-x)/\sigma$. By splitting the integral appearing in the last line of Eqs.~(\ref{eq:DeltaRstep1}),
\begin{align}
\Delta R_{ud}^{(\tau)}(\sigma) 
&=
-\sigma^3
\left\{
\int_{0}^{\frac{1}{\sigma}}\dd y\,\left[
1 -\Theta_1\left(y\right)\right]\,
y^2\Tilde{\rho}(1-\sigma y)
-
\int_{-\infty}^{0}\dd y\,
\Theta_1\left(y\right)\,
y^2\Tilde{\rho}(1-\sigma y)
\right\}\;,
\end{align}
and by relying again on Eqs.~(\ref{eq:thetaprops}), we have
\begin{align}
\Delta R_{ud}^{(\tau)}(\sigma) 
&=
-\sigma^3
\left\{
\int_{0}^{\frac{1}{\sigma}}\dd y\,\left[
1 -\Theta_1\left(y\right)\right]\,
y^2\Tilde{\rho}(1-\sigma y)
-
\int_{0}^{\infty}\dd y\,
\Theta_1\left(-y\right)\,
y^2\Tilde{\rho}(1+\sigma y)
\right\}
\nonumber \\[8pt]
&=
-\sigma^3
\left\{
\int_{0}^{\infty}\dd y\,\left[
1 -\Theta_1\left(y\right)\right]\,
y^2\Tilde{\rho}(1-\sigma y)
-
\int_{0}^{\infty}\dd y\,
\Theta_1\left(-y\right)\,
y^2\Tilde{\rho}(1+\sigma y)
\right\}
+\order{e^{-\frac{1}{\sigma}}}
\nonumber \\[8pt]
&=
\sigma^3
\int_{0}^{\infty}\dd y\,\left[
1 -\Theta_1\left(y\right)\right]\,
y^2
\left[
\Tilde{\rho}(1+\sigma y)
-
\Tilde{\rho}(1-\sigma y)
\right]
+\order{e^{-\frac{1}{\sigma}}}\;.
\label{eq:DeltaRstep2}
\end{align}
In deriving the previous result we only assumed that $\Tilde{\rho}(x)$ is a tempered distribution and therefore that, as such, it grows at most as a power in the $x\mapsto \infty$ limit. In order to make further progress we now need to know the behaviour of $\Tilde{\rho}(x)$ around $x=1$. Indeed, as we are now going to show, the behaviour of $\Delta R_{ud}^{(\tau)}(\sigma)$ w.r.t.\ $\sigma$ is strongly dependent upon the behaviour of $\Tilde{\rho}(x)$ around $x=1$.

Since, axiomatically, $\Tilde{\rho}(x)$ is a tempered distribution, we cannot exclude a singular behaviour at $x=1$. We therefore consider the following rather general decomposition
\begin{align}
	\Tilde{\rho}(x)=\Tilde{\rho}_\mathrm{reg}(x) + \Tilde{\rho}_+(x)\theta(1-x) + \sum_{d=0}^{N_d}\Tilde{\rho}_d\,\partial^d_x\,\delta(1-x)\,,
 \label{eq:rhotildedec}
\end{align}
where we assume that $\Tilde{\rho}_{\text{reg}}(x)$ and $\Tilde{\rho}_{+}(x)$ are $\mathcal{C}_\infty$ regular functions at $x=1$.
The contributions proportional to the $\theta(1-x)$ and to the derivatives of $\delta(1-x)$ have been introduced to parametrize possible discontinuities and $\delta$-function singularities at $x=1$. We are now going to analyze in turn the three contributions to $\Delta R_{ud}^{(\tau)}(\sigma)$, 
\begin{align}
	\Delta R_{ud}^{(\tau)}(\sigma)=
    \Delta R_\mathrm{reg}(\sigma) +     
    \Delta R_+(\sigma) +
    \Delta R_\delta(\sigma)\;, 
    \label{eq:DeltaRsplit}
\end{align}
corresponding to the decomposition of $\Tilde{\rho}(x)$ given in Eq.~(\ref{eq:rhotildedec}).

The asymptotic expansion of $\Delta R_\mathrm{reg}(\sigma)$ can readily be obtained by substituting in Eq.~(\ref{eq:DeltaRstep2}) the Taylor series expansion
\begin{align}
\Tilde{\rho}_\mathrm{reg}(1+\sigma y)
-
\Tilde{\rho}_\mathrm{reg}(1-\sigma y)
=
\sum_{n=0}^\infty
\frac{2(\sigma y)^{2n+1}}{(2n+1)!}
\Tilde{\rho}_\mathrm{reg}^{(2n+1)}(1)
\end{align}
and by defining the numeric coefficients
\begin{align}
C_\Theta^{n}
=
\int_{0}^{\infty}\dd y\,\left[
1 -\Theta_1\left(y\right)\right]\,
y^n \;.
\end{align}
We have
\begin{align}
\Delta R_\mathrm{reg}(\sigma)
=
\sigma^4
\sum_{n=0}^\infty
\frac{2(\sigma)^{2n}}{(2n+1)!}
\Tilde{\rho}_\mathrm{reg}^{(2n+1)}(1)
\,
C_\Theta^{2n+3}
=\order{\sigma^4}
\;.
\label{eq:DeltaRreg}
\end{align}
Therefore, under the assumption that $\rho_{\mathrm{L}}(E/m_\tau)$ and $\rho_{\mathrm{T}}(E/m_\tau)$ are both regular at $E=m_\tau$, we have thus derived the result stated in Eq.~(\ref{eq:sigmascaling}) and used in the main text to fit $R_{ud}^{(\tau,I)}(\sigma)$ according to Eq.~(\ref{eq:sigma4_ansatz}).

We now discuss the singular contributions $\Delta R_+(\sigma)$ and $\Delta R_\delta(\sigma)$. The asymptotic expansion of $\Delta R_+(\sigma)$ can be derived by 
first noticing that the term proportional to $\theta(1-(1+\sigma y))$ gives no contribution to the integral of Eq.~(\ref{eq:DeltaRstep2}) and then by substituting in that expression the expansion
\begin{align}
\Tilde{\rho}_+(1-\sigma y)
=
\sum_{n=0}^\infty
\frac{(-\sigma y)^{n}}{n!}
\Tilde{\rho}_+^{(n)}(1)\;.
\end{align}
We get
\begin{align}
\Delta R_+(\sigma)
=
-
\sigma^3
\sum_{n=0}^\infty
\frac{(-\sigma)^{n}}{(n)!}
\Tilde{\rho}_+^{(n)}(1)
\,
C_\Theta^{n+2}
=\order{\sigma^3}
\;.
\label{eq:DeltaRtheta}
\end{align}

In order to derive the asymptotic expansion of $\Delta R_\delta(\sigma)$ it is convenient to start from the representation of $\Delta R_{ud}^{(\tau)}(\sigma)$ given in Eq.~(\ref{eq:DeltaRstep1}). We have
\begin{align}
\Delta R_\delta(\sigma) 
&=
\sigma^3\sum_{d=0}^{N_d}
\Tilde{\rho}_d
\int_{-\infty}^{\frac{1}{\sigma}}\dd y\,\left[
\Theta_1\left(y\right)
-
\theta(y)
\right]\,
y^2\,
\frac{\partial^d}{\partial^d(1-\sigma y)}
\delta(\sigma y)
\nonumber \\[8pt]
&=
\sigma^2\sum_{d=0}^{N_d}
(-1)^d
\sigma^{-d}
\Tilde{\rho}_d
\int_{-\infty}^{\frac{1}{\sigma}}\dd y\,\left[
\Theta_1\left(y\right)
-
\theta(y)
\right]\,
y^2\,
\partial_y^d
\delta(y)
\nonumber \\[8pt]
&=
\sigma^2\sum_{d=0}^{N_d}
\sigma^{-d}
\Tilde{\rho}_d
\partial_y^d
\left\{
y^2\,
\left[
\Theta_1\left(y\right)
-
\theta(y)
\right]
\right\}_{y=0}\;.
\label{eq:DeltaRdelta}
\end{align}

The results derived in this appendix deserve some remarks. We presented a rather general analysis of the asymptotic behaviour of $\Delta R_{ud}^{(\tau)}(\sigma)$ for small values of $\sigma$. Our results have been obtained by parametrizing the possible singularities of the tempered distributions $\rho_{\mathrm{L}}(E/m_\tau)$ and/or $\rho_{\mathrm{T}}(E/m_\tau)$ according to Eq.~(\ref{eq:rhotildedec}). Correspondingly, see Eq.~(\ref{eq:DeltaRsplit}), we studied the contribution to the asymptotic behaviour of $\Delta R_{ud}^{(\tau)}(\sigma)$ induced by the regular part of the spectral densities ($\Delta R_\mathrm{reg}(\sigma)$) and by the possible occurrence of discontinuities ($\Delta R_+(\sigma)$) or $\delta$-function singularities ($\Delta R_\delta(\sigma)$). This has been done because, from the axiomatic viewpoint, we cannot exclude singularities of the hadronic spectral densities at $E=m_\tau$. On the other hand, from the physical viewpoint, it is extremely unlikely that the infinite volume hadronic spectral densities are singular at a value of the energy corresponding to a leptonic scale, i.e.\ at $E=m_\tau$. 

Unphysical singularities at $E=m_\tau$ might accidentally be observed at finite volume. This might happen if one considers too-small values of the smearing parameter $\sigma$ before performing the infinite volume limit of the lattice data. In this case, according to Eq.~(\ref{eq:DeltaRtheta}) and Eq.~(\ref{eq:DeltaRdelta}), one might even see a singular behaviour of $\Delta R_{ud}^{(\tau)}(\sigma)$ for (too) small values of $\sigma$.

In absence of singularities our result of Eq.~(\ref{eq:DeltaRreg}) shows that only even powers of $\sigma$ appear in the asymptotic expansion of $\Delta R_{ud}^{(\tau)}(\sigma)$ and that the leading behaviour is $\Delta R_{ud}^{(\tau)}(\sigma)=\order{\sigma^4}$. The numerical results presented in the main text are fully compatible with the expected regular behaviour (see Figure~\ref{fig:sigma_dep}) and this, in the light of the previous observations, can be seen as a reassuring evidence concerning the fact that we performed the infinite volume limit of our lattice data by estimating reliably the associated systematic errors.

\bibliography{biblio}
\bibliographystyle{JHEP}

\end{document}